# Guidelines for band gap opening in graphene superlattices with periodic π-vacancy distribution

Diyan Unmu Dzujah,[1] Hongde Yu,[1] and Thomas Heine[1, 2, 3*]

[1] Faculty of Chemistry and Food Chemistry, Technische Universität Dresden, 01062 Dresden, Germany

[2] Center for Advanced System Understanding CASUS, Helmholtz-Zentrum Dresden-Rossendorf e. V., 01328 Dresden, Germany

[3] Department of Chemistry, Yonsei University, Seodaemun-gu, 03722 Seoul, Republic of Korea

**ABSTRACT**. Periodic π-vacancies in graphene superlattices (GSLs) provide a symmetry-based route to band gap opening in graphene by modifying the π-band dispersion. Yet, the symmetry conditions that determine whether a vacancy motif can open a band gap remain unclear. Here, we investigate periodic π-vacancy GSLs using a nearest-neighbor tight-binding (TB) model with one $p_z$ orbital per carbon site to identify the symmetry requirements for band gap opening. π-vacancies, resembling functionalized, substituted, or missing carbon sites, are modeled as site deletions in the $\pi$ basis, and the corresponding hopping matrix elements to and from the deleted sites are set to zero. We focus on π-vacancy motifs with $C_2$ and $C_3$ point groups. A $3n \times 3n$ GSL, where $n = 1, 2, 3, \ldots$ is the integer scaling factor multiplying the honeycomb primitive-cell vectors, is required to fold K and K′ to Γ, which is one of the requirements for opening a band gap. For $C_3$-type vacancies, the Dirac cones are pinned at the high-symmetry points, thus remaining at Γ in the folded $3n$ GSLs. However, those $C_2$-type vacancies which reduce the global point group of the GSL to $D_{2h}$ by preserving a pair of perpendicular mirror symmetries ($\sigma_v \perp \sigma_d$), can constrain the positions of Dirac cones at Γ. The absence of the $\sigma_v$ and $\sigma_d$ planes allows the cones to shift around Γ ($\pm\Delta\mathbf{q}$, $\pm\Delta\mathbf{q}$) in the $3n$ superlattice**.** The $C_3$-type vacancies open a relatively larger band gap compared to the $C_2$-type vacancies due to twofold degeneracy at Γ, while the $C_2$-type vacancies leave the GSL with nondegenerate bands. The energy contours near the Dirac cone for $C_3$-symmetric vacancies remain isotropic, while the absence of $C_3$ generates anisotropic cones. The band gap openings in $3n$ superlattices scale linearly with defect concentration for each vacancy motif, with the $D_{6h}$ motif yielding the largest gap, reaching 314 meV at only 3.7% defect concentration. Disorder of the superlattice by missing vacancies or by vacancy displacements of one or two primitive lattice vectors of graphene breaks the $C_3$ symmetry, lift the degeneracy, and reduce the gap toward the pristine graphene limit.

## **I. INTRODUCTION**.

Graphene exhibits symmetry-protected Dirac cones at K and K′and is intrinsically gapless in the absence of spin-orbit coupling [1]. Dirac cones yield linear band dispersion near the inequivalent K and K′ valleys, allowing electrons to behave as Dirac quasiparticles with Fermi velocity $v_F \approx 10^6 m/s$, giving rise to high carrier mobility and unusual transport properties [2,3,4]. Despite these electronic properties, the absence of a band gap in monolayer graphene limits its application in electronic devices that require one, such as field-effect transistors. This limitation has motivated extensive work on band gap engineering in graphene, including chemical doping [5,6,7], vacancies [8,9,10], functionalization [11,12,13], and Kekulé-type bond modulation [14]. Among these approaches, periodic π-vacancies are a particularly effective approach to defect engineering because they directly modify the π-electron network and disrupt the bipartite hopping connectivity. Within the tight-binding (TB) framework, the relevant defects are often $\pi$-vacancy patterns involving several neighboring sites rather than single-site vacancies [9,10]. The $\pi$-vacancies are removed sites in the TB Hamiltonian which reflect local functionalization, e.g. by BN substitution [5,6,7], or by forming additional bonds leading to transformation into $sp^3$ sites [12,13,15]. From this perspective, a $\pi$-orbital TB description offers a direct framework that captures the essential physics of gap opening, while remaining computationally efficient for large GSLs.

Graphene defects can be introduced experimentally either in an aperiodic or periodic manner. Aperiodic defects can be introduced through top-down processes, such as electron/ion irradiation, which create vacancies by knocking out carbon atoms [16,17], or plasma etching [18,19]. These methods are valued for their scalability up to the wafer scale, however, the lack of atomic precision and uncontrolled defect geometry can affect the reproducibility of band gap opening [16,17,19]. In contrast, bottom-up approaches can yield well-ordered periodic defect arrays through molecular assembly, providing near-atomic-level precision in defect placement [20,21,22]. In defected graphene, the band gap depends on the shape, size, and arrangement of the defect patterns [18]. However, realistic structures still exhibit motif displacement [19,20,23,24], which alters the effective symmetry of the perturbation and thereby modifies the electronic properties, including band gap suppression.

Introducing a periodic pattern of vacancies creates a periodic superlattice modulation in the system [25,26]. A superlattice is defined by a real-space periodicity larger than graphene's primitive cell, thereby reducing the Brillouin zone (BZ) to a mini-Brillouin zone (mBZ) and backfolding the pristine bands. Previously studied examples include graphene antidot lattices, also known as nanomesh graphene, and related periodic porous carbon networks that modulate the periodic potential in graphene superlattices [9,27,28]. Prior work shows that the band dispersions of graphene superlattices are symmetry-dependent [10,29,30], and identifying the relevant symmetries provides a practical route to gap opening in graphene superlattices (GSLs). In the TB model, the Dirac cone in pristine graphene is protected by time-reversal and inversion symmetries, which map K and K′, together with sublattice symmetry, which enforces a symmetric spectrum [31], with the global point group of graphene being $D_{6h}$. The cones at $K_1$, $K_2$, $K_3$ in the BZ of pristine graphene are related by a $C_3$ rotation, as are $K'_1$, $K'_2$, $K'_3$, and $C_2$ swaps K and K′, as shown in Fig. 1(a). Theoretical studies of graphene antidot lattices using TB have predicted that a band gap can be opened in the $3n \times 3n$ superlattice, yielding a fourfold degeneracy at the Γ point [32,33,34]. This is a direct consequence of the translational symmetry $T_R: \mathbf{r} \rightarrow \mathbf{r} + \mathbf{R}$ in graphene, which determines the folding positions of the Dirac cones across different supercell sizes, interchanges the folded locations of K and K′ between the $3n + 1$ and $3n - 1$ supercells, and pins both K and K′ at Γ in the 3n supercells, as shown in Fig. 1(b) and (c). Therefore, the symmetry constraints acting in the folded low-energy subspace can be used to formulate design rules for vacancy superlattices.

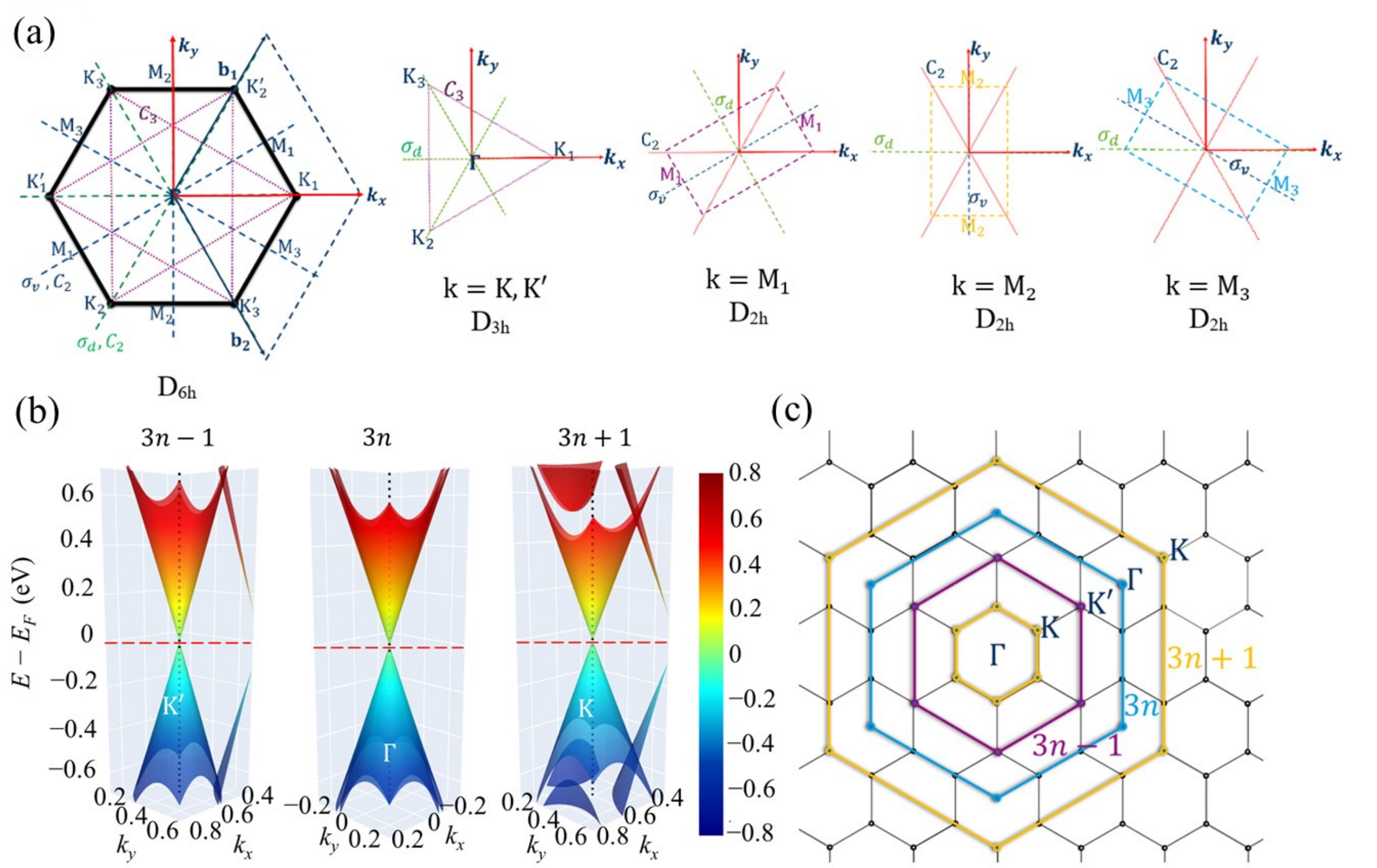

FIG. 1 (a) Graphene Brillouin zone showing reciprocal lattice vectors $\mathbf{b_1}$ and $\mathbf{b_2}$, the global point group $D_{6h}$, and the little group at each high-symmetry point $(\Gamma, \mathrm{K}, \mathrm{K}', \mathrm{M}_1, \mathrm{M}_2, \text{and } \mathrm{M}_3)$. Representative mirror symmetries across $\Gamma - M$ $\sigma_v: (k_x, k_y) \to (-k_x, k_y)$ and $\Gamma - \mathrm{K}$ $\sigma_d: (k_x,\ k_y) \to (k_x, -k_y)$ are repeated every 120°, resulting in three mirror lines for each of $\sigma_v$ and $\sigma_d$. The $C_2$ and three $\sigma_v$ planes interchange K and K′, while the three $\sigma_d$ planes connect $\mathrm{K}_1 \to \mathrm{K}_1$ and $\mathrm{K}_2 \leftrightarrow\ \mathrm{K}_3$. Combining $C_3$ with three $\sigma_d$ or $\sigma_v$ planes give the little group $D_{3h}$ at K and K′. The $\mathrm{M}_1$, $\mathrm{M}_2$, and $\mathrm{M}_3$ points are the midpoints of the BZ edges and have the little group $D_{2h}$. (b) Dirac point locations for all supercell sizes of pristine graphene. (c) Schematic band folding mechanism in pristine graphene, illustrating how valleys are mapped into the reduced zone.

In order to make this explicit, within the $\pi$-orbital TB model, graphene is described as a bipartite lattice with sublattices A and B, where the Bloch Hamiltonian can be written as:

$$H(\mathbf{k}) = \begin{bmatrix} \varepsilon_{pz} & f(\mathbf{k}) \\ f^*(\mathbf{k}) & \varepsilon_{pz} \end{bmatrix} \tag{1}$$

where $\varepsilon_{pz} = 0$ and $f(\mathbf{k})$ is the structure factor encoding the nearest-neighbor hoppings, $f(\mathbf{k}) = -t\sum_{n=1}^{3} e^{i\mathbf{k}\cdot\boldsymbol{\delta}_\mathrm{n}}$, $t$ is the nearest-neighbor hopping amplitude and $\boldsymbol{\delta}_\mathrm{n}$ are the nearest-neighbor vectors connecting A and B sites, and the dispersion energy is then $E_\pm(\mathbf{k}) = \pm|f(\mathbf{k})|$. The Dirac points arise when the sum of the Fourier components vanishes, which occurs at K and K′ where $f(\mathrm{K}) = f(\mathrm{K}') = 0$. A first-order Taylor expansion about the K, with $\mathbf{k} = \mathbf{K} + \mathbf{q}$ and $|\mathbf{q}| \ll |\mathbf{K}|$, yields a dispersion linear in $|\mathbf{q}|$ [2]. The formulation shows that periodic modification of the π-network in graphene can be viewed as a modulation of the structure factor, which controls the low-energy dispersion of the GSL.

Previous studies have predominantly focused on highly symmetric defect motifs and superlattices [9, 10, 27, 30], whereas lower-symmetry configurations, which may arise from structural imperfections or experimental disorder, have received comparatively less attention. In this work, we establish symmetry-based guidelines for band gap opening in periodic $\pi$-vacancy GSLs using a TB framework that captures the low-energy physics associated with the missing $\pi$ states. We determine how point-group constraints govern symmetry-allowed gap formation and how these constraints depend on the $3n$ superlattice commensurability condition, which is a prerequisite for band gap opening. Additionally, we investigate how vacancy displacement modifies the effective symmetry of the superlattice. We then assess how this symmetry reduction impacts the robustness and magnitude of the band gap. We find that robust and substantial gap opening in $3n$ superlattices occurs when $C_3$ symmetry is preserved. In contrast, a smaller band gap is obtained when the sufficient symmetry condition of two orthogonal mirror planes remains satisfied. Otherwise, gap opening is not guaranteed. Finally, we show how motif displacement lowers the effective symmetry and suppresses the gap by lifting the twofold degeneracies.

## II. METHODS

The GSLs with periodic vacancies are modeled using the TB method implemented in the PythTB package [35]. This study neglects spin polarization and spin-orbit coupling, so the dispersion is spin-independent and time-reversal symmetry is preserved. The band structures of the GSLs are treated within the nearest-neighbor approximation, with $t = -2.7$ eV and the on-site energy $\varepsilon_i = 0$. The TB Hamiltonian is defined as:

$$H = \sum_i \varepsilon_i c_i^\dagger c_i + \sum_{\langle i,j \rangle} t\, c_i^\dagger c_j \quad (2)$$

where $c_i^\dagger$ is the creation operator, $c_j$ is the annihilation operator for the $p_z$ orbital on site $i$, and $\langle i,j \rangle$ denotes the nearest-neighbor pairs. The vacancy periodicity is set by an $N \times N$ superlattice, where $N = 3n - 1,\ 3n,$ or $3n + 1$, and $n = 1, 2, 3, \ldots$, is the integer scaling factor of the honeycomb primitive-cell vectors $\mathbf{a_1}$ and $\mathbf{a_2}$, as shown in Fig. 2(a), and the graphene lattice constant is defined as $a = |\mathbf{a_1}| = |\mathbf{a_2}|$. The vacancies are modeled by removing the $p_z$ orbital of C atoms in the GSL. We focus on $C_3$-type vacancies ($C_{3h(12C)}$, $D_{3h(A\neq B)}$, $D_{3h(18C)}$, $D_{6h(6C)}$) and $C_2$-type vacancies ($C_{2h(4C)}$, $D_{2h(2C)}$, $D_{2h(10C)}$, $D_{2h(16C)}$) as shown in Fig. 2(b).

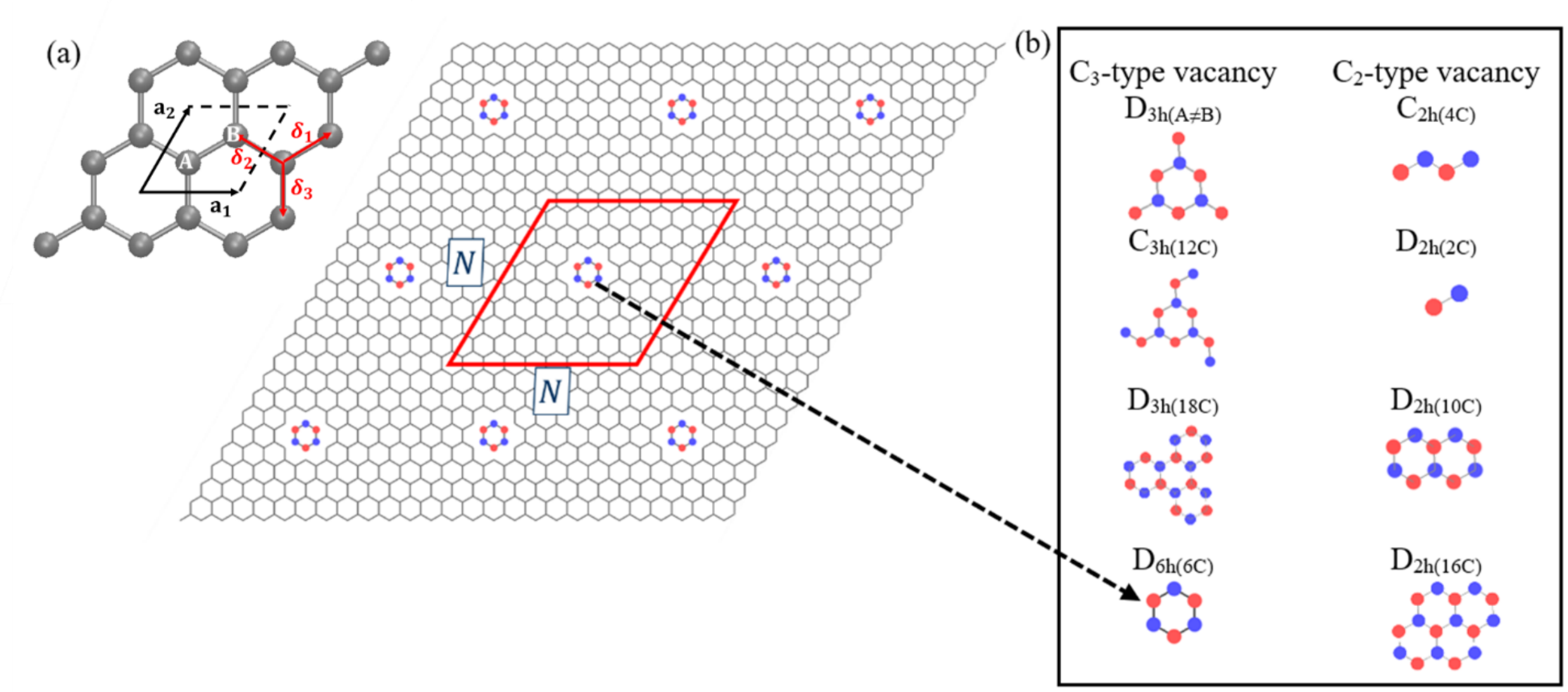

FIG. 2. (a) $N \times N$ GSL defined by the integer scaling factor $n$ applied to the primitive lattice vectors $\mathbf{a_1}$ and $\mathbf{a_2}$. Red (blue) markers denote removed sites on the A (B) sublattices used to model the $\pi$-vacancies. (b) $C_2$- and $C_3$-type $\pi$-vacancies considered in this work.

## III. RESULTS AND DISCUSSIONS

Different π-vacancy types in the GSL induce different effects in the band structure. In general, we observe three effects upon introducing a periodic vacancy to the GSL, namely (i) zero modes, where the eigenvalues remain at zero energy across momentum space and appear as a flat band, (ii) the shift of the Dirac cones towards low-symmetry points in the BZ, and (iii) band gap opening. Vacancies with $\mathrm{A} \neq \mathrm{B}$ generate at least $m$ zero-energy modes, where $m = |\mathrm{A} - \mathrm{B}|$, due to the sublattice imbalance, consistent with Lieb's theorem [36]. For example, $D_{3h}$ vacancy shows flat bands as shown in Table I of the Supplemental Material [37], and may induce magnetic properties [29,36,38,39]. Therefore, in this study we focus on cases with $\mathrm{A} = \mathrm{B}$, which is required for band gap opening in the GSL. The effects (ii) and (iii) are discussed in the following sections.

### A. Position of the Dirac cone

The position of the Dirac cones in momentum space is dictated by the symmetry of the structure. $C_3$ symmetry enforces that the two valleys K and K′ are 120° rotationally equivalent. Loss of $C_3$ symmetry in the GSL can shift the position of the Dirac cone in reciprocal space by $\Delta\mathbf{q}$, or, if certain symmetry requirements are met, may lead to band gap opening. It is important to note that the Dirac points can shift to low-symmetry points that lie off the high-symmetry pathways and therefore do not appear in common band structure plots. Therefore, plotting the 2D band structure is essential to visually observe the effect of this perturbation, as shown in Fig. S1 of the Supplemental Material [37]. The 2D band structures of pristine graphene are depicted in Fig. 3(a), showing that Dirac cones are located at the high-symmetry points in each superlattice. The introduction of the $C_3$-type vacancy in the GSL keeps the Dirac cones at high-symmetry points, as shown in Fig. 3(b), thus, any band gap opening can be observed by sampling these points. In contrast, the absence of the $C_3$ axis can shift the Dirac cones to $K \pm \Delta\mathbf{q}$ in the $3n-1$ and $3n+1$ superlattices, as depicted in Figs. 3(c-d). The magnitude of $\Delta\mathbf{q} \propto$ defect concentration, meaning a small perturbation of the $C_2$-type vacancy leads to minimal $\Delta\mathbf{q}$.

### B. Band gap opening in the GSL and the effect of the $C_3$ and $\sigma_d \perp \sigma_v$

A necessary condition for the band gap opening in a graphene superlattice is that both K and K′ are folded onto the Time-Reversal Invariant Momenta (TRIM) point Γ [33], which occurs when the superlattice size is $3n$. The $C_3$-type vacancy in the GSL makes the folding straightforward due to the absence of $\Delta\mathbf{q}$, keeping the Dirac cone at a high-symmetry point. Fig. 3(b) shows the positions of the Dirac cones in $3n-1,\ 3n$, and $3n+1$ for $n\ =\ 3$ with a $D_{3h(18C)}$ vacancy. A band gap of $|\Delta_g| = 615$ meV is observed at Γ in the 3n superlattice. The previously observed fourfold degeneracy at $E_F$ is split into a twofold degeneracy, with the global point group of the mBZ being $D_{3h}$. The little group is $D_{3h}$ at K and K′, and $D_{2h}$ at the M points. In the GSL containing $D_{3h(18C)}$ vacancy, the identity, time-reversal, inversion, $C_2$, $C_3$, $3\sigma_v, 3\sigma_d,$ and $\sigma_h$ symmetries are well preserved, as illustrated in the Fig. 4 (a). The periodic vacancy in GSL with sublattice removal A=B keeps the sublattice symmetry and preserve the bipartite form of the Hamiltonian matrix. This shows that the matrix chirality $\Gamma H \Gamma^{-1} = -H,\ \Gamma^2 = I$ is preserved, leading to symmetric eigenvalues around $E_F$ in all GSL sizes. The $E(k) = -E(k)$ at the Conduction Band Maximum (CBM) and Valence Band Minimum (VBM) in the $3n$ GSL can be seen in Fig. 3. As a result, a band gap opens.

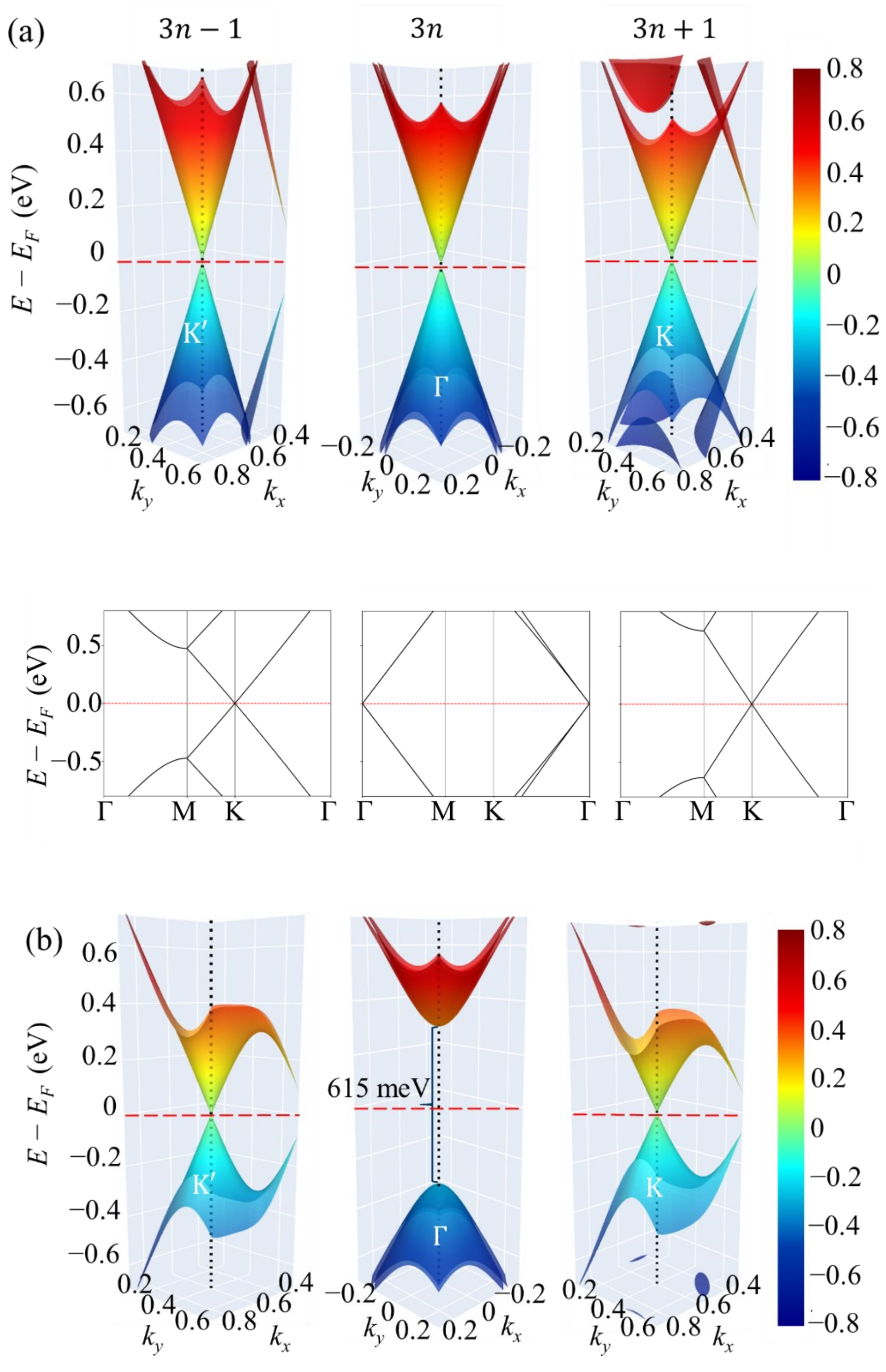

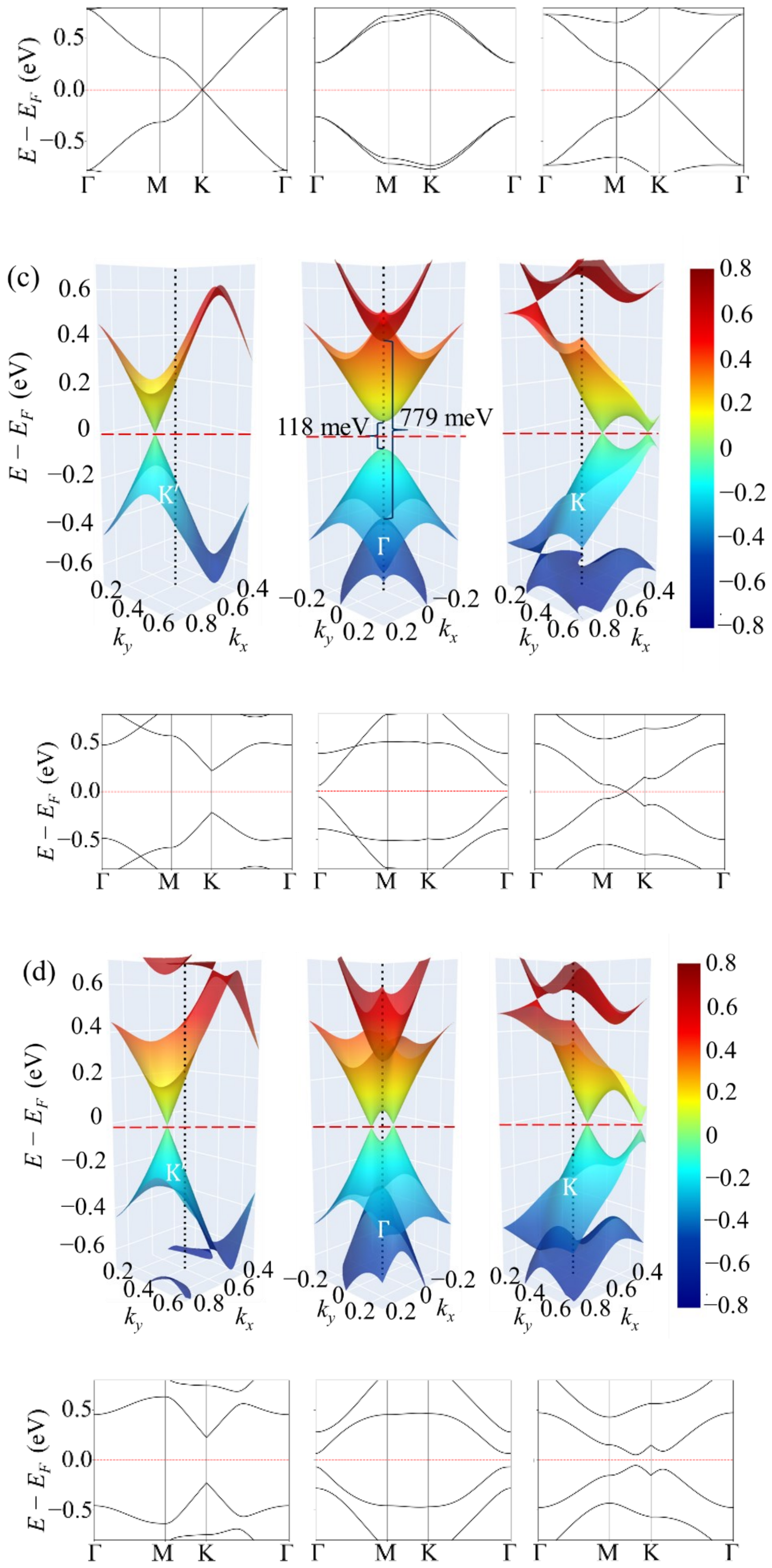
(c)
118 meV
779 meV
K′
Γ
K
(d)
K
Γ
K
$E - E_F$ (eV)
Γ M K Γ
$k_y$
$k_x$

FIG. 3. 2D (top) and 1D (bottom) band structure of (a) pristine graphene, GSLs with (b) $D_{3h(18C)}$, (c) $D_{2h(16C)}$, and (d) $C_{2h(4C)}$ vacancies for superlattice sizes $3n-1$, $3n$, and $3n+1$, where $n=3$, $E_F$ is denoted by the red dash. The band structures of other vacancy motifs can be found in Table I of the Supplemental Material [37].

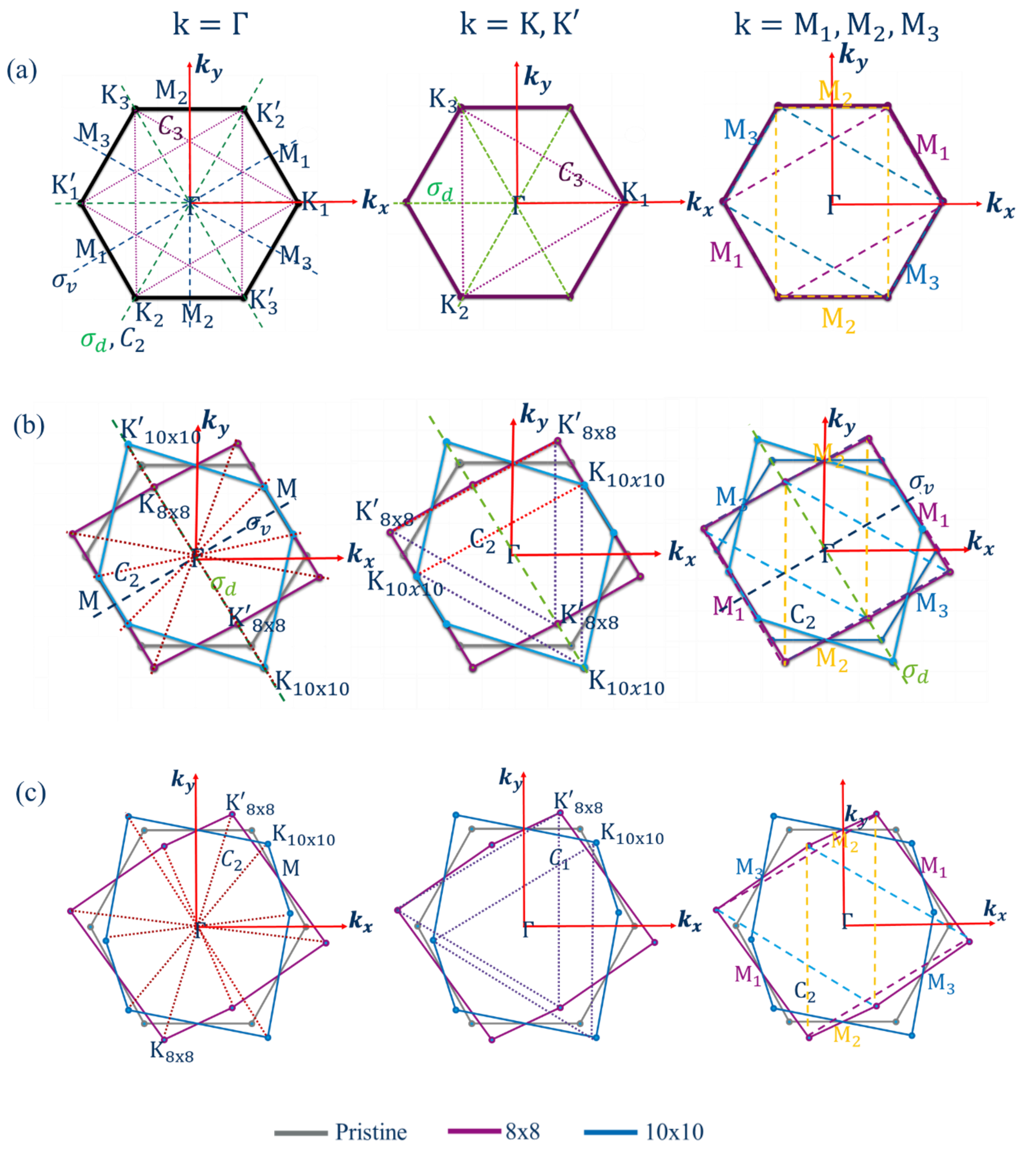

FIG. 4. Reduced little groups at the high-symmetry points ($\Gamma, \mathrm{K}, \mathrm{K}', \mathrm{M}_1, \mathrm{M}_2$, and $\mathrm{M}_3$) in the GSL with (a) $D_{3h(18C)}$, (b) $D_{2h(16C)}$, and (c) $C_{2(4C)}$ vacancies for superlattice sizes 8×8 and 10×10.

In contrast, whether $C_2$-type vacancies without a $C_3$ rotational axis open a band gap depends on which symmetries remain in the mBZ and whether the superlattice periodicity permits momentum space commensurability. Figs. 3(b) and 4(b) show the positions of the Dirac cones in the GSL with a $D_{2h(16C)}$ vacancy slightly shifted off the high-symmetry point due to the breaking of $C_3$ symmetry, leading to a reduction of the little group at K and $\mathrm{K}'$ to $C_{2v}$ = (E, $C_2$, $\sigma_d, \sigma_h$), with the little group at $\mathrm{M}_1$ remaining $D_{2h}$, while $\mathrm{M}_2$ and $\mathrm{M}_3$ are reduced to $C_{2h}$. A direct consequence of $C_3$ breaking is the loss of at least two $\sigma_d$ and two $\sigma_v$ mirror planes, because the product of $\sigma^{0°}\sigma^{60°} = C_{120°} = C_3$, which is not possible after losing the $C_3$ axis, reducing the point group of the mBZ to $D_{2h}$. In this case, despite $\Delta\mathbf{q} \neq 0$ in the $3n-1$ and $3n+1$ GSLs, in the $3n$ GSLs both K and $\mathrm{K}'$ can still fold to $\Gamma$ and open a band gap of 118 meV, as seen in Fig. 3(c). The $D_{2h(16C)}$ GSL preserves one pair of $\sigma_d \perp \sigma_v$ mirror planes, with one $\sigma_d$ constraining K and $\mathrm{K}'$ to shift only along the mirror line $k_{\parallel}$, while shifts perpendicular to the mirror line $k_{\perp}$ are forbidden, reducing the global point group to $D_{2h}$ as shown in Fig. 4(b). Additionally, the constraints imposed by the $\sigma_d \perp \sigma_v$ planes lock the K and $\mathrm{K}'$ at $\mathrm{k} = \Gamma = (0,0)$ in the $3n$ GSL because both mirrors forbid the shift along $k_{\perp}$ perpendicular to each other. Among the $C_2$-type vacancies studied in this work, the $D_{2h(2C)}$, $D_{2h(10C)}$, and $D_{2h(16C)}$ vacancies preserve one pair of $\sigma_d \perp \sigma_v$ planes as shown in Fig. 4(b) and Fig. S2 of the Supplemental Material [37]. These $C_2$-type vacancies preserve identity, time-reversal, inversion, $C_2$, $\sigma_v$, $\sigma_d$, and $\sigma_h$ symmetries. The $C_{2h(4C)}$ vacancy breaks all three $\sigma_v$ and $\sigma_d$ mirror planes in the mBZ, leaving no constraint to pin K and $\mathrm{K}'$ during band folding to $\Gamma$ in the $3n$ GSL, allowing the shift $(\pm\Delta\mathbf{q}, \pm\Delta\mathbf{q})$ around $\Gamma$, preventing the folding of K and $\mathrm{K}'$ to $\Gamma$ in the $3n$ GSL and reducing the point group of the mBZ, and the little groups at $M_1$, $M_2$, and $M_3$ to $C_{2h}$, while the little groups at K and K′ are $C_s$ as shown in the Fig. 4(c). These types of GSLs preserve identity, time-reversal, inversion, $C_2$, and $\sigma_h$ symmetries and keep the GSL gapless as shown in Fig. 3(d). Thus, the sufficient symmetries in the mBZ for $C_2$-type vacancies to enforce the locking of K and K′ to Γ in $3n$ GSLs are a pair of perpendicular mirror planes, $\sigma_d \perp \sigma_v$.

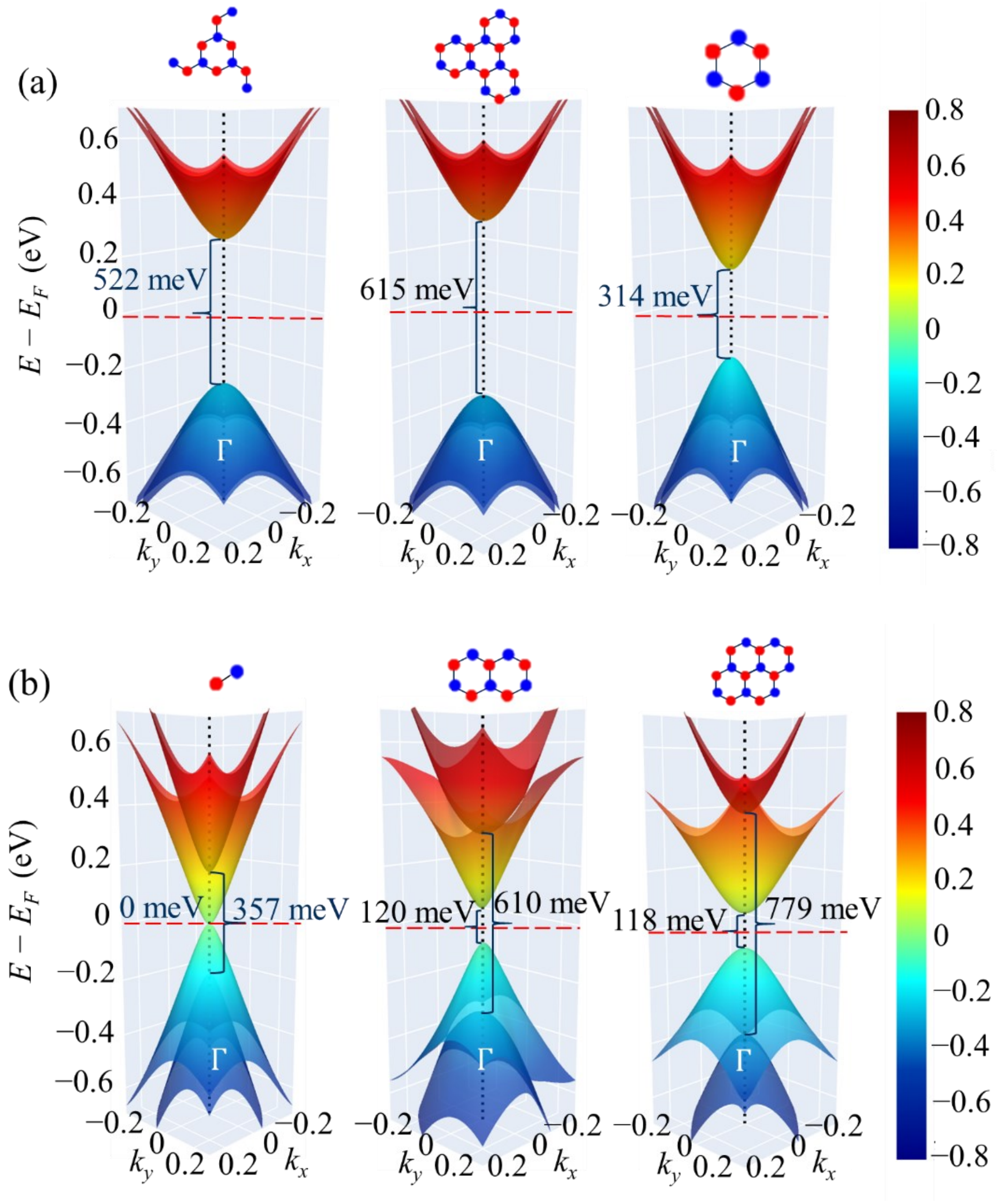

FIG. 5. Band structures $3n$ GSLs with (a) $C_3$-type vacancies, and (b) $C_2$-type vacancies, with $n = 3$, illustrating the influence of symmetry on band gap opening around Γ.

Band gaps obtained from the $C_2$-type vacancies are smaller than those from the $C_3$-type vacancies. This is a consequence of breaking the doubly degenerate states at $\pm\frac{1}{2}|\Delta E_1|$ into nondegenerate states near $E_F$ as shown in Figs. 5 and 6, indicating the loss of $C_3$ symmetry in the GSL. A higher defect concentration or a smaller superlattice size leads to a larger band gap in the GSL. Here, the defect concentration is defined as the percentage of removed $\pi$ orbitals relative to the total number of $\pi$ orbitals in the pristine graphene $N \times N$ supercell. With decreasing perturbation, the characteristics of the GSL become closer to those of pristine graphene, this is depicted in Figs. 6(a, b). Fig. 6(b) shows that the band gap opening in $3n$ superlattices follows a linear trend for each vacancy motif, as summarized in Table III of the Supplemental Material [37], with $D_{6h(6C)}$ yielding the widest band gap as a function of defect concentration, reaching 314 meV at a defect concentration of only 3.70%. By comparison, $C_{3h(12C)}$ and $D_{3h(18C)}$ require defect concentrations of 4.17% and 6.25% to reach 295 meV and 345 meV, respectively. In contrast, smaller band gaps are obtained for $C_2$-type vacancies. $D_{2h(10C)}$ requires a defect concentration of 6.17% to reach 120 meV, whereas $D_{2h(16C)}$ requires 9.88% to reach 118 meV, and $D_{2h(2C)}$ cannot open a band gap. In the 9×9 GSL, $D_{6h(6C)}$ gives a band gap 2.6 times larger than that of $D_{2h(10C)}$, making $D_{6h(6C)}$ the most efficient periodic vacancy motif for opening a band gap in the GSL. Therefore, the vacancy motifs can be ordered by increasing required defect concentration as $D_{6h(6C)} < C_{3h(12C)} < D_{3h(18C)} \ll D_{2h(10C)} < D_{2h(16C)}$. These results show that preserving $C_3$ symmetry is necessary to open a large band gap in $3n$ GSLs with periodic $\pi$-vacancies. The $C_2$-type vacancies show band-splitting of two degenerate bands at VBM and CBM ($\Delta E_2$), and the magnitude is comparable to the energy scale of the degenerate bands in $C_3$-type vacancies. This implies that the K and K′ are unequally split, lowering the magnitude of the band gap $\Delta E_1$ as shown in Figs. 6(c, d).

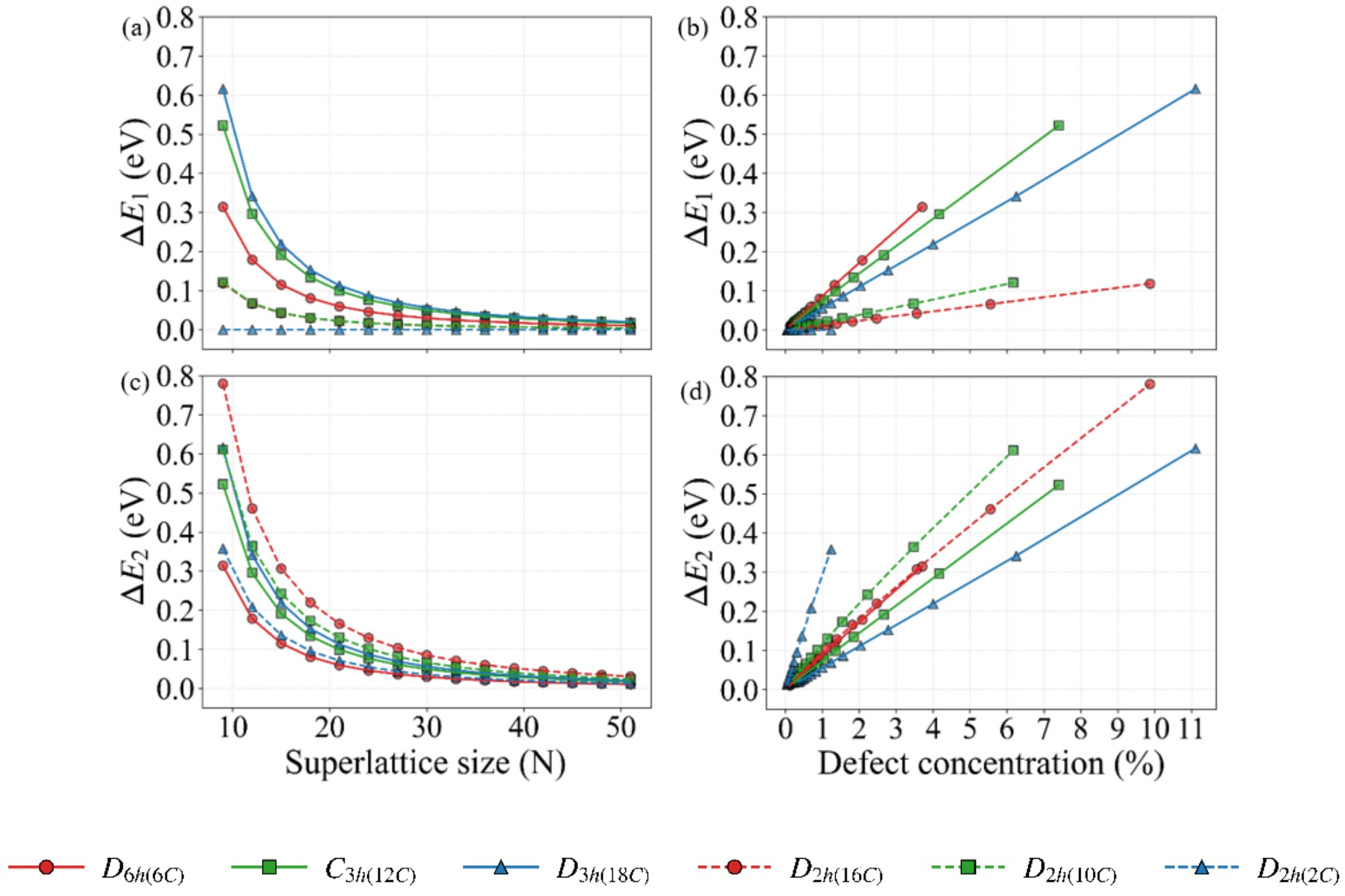

FIG. 6. Band gap $\Delta E_1$ as a function of (a) superlattice size $N$ and (b) defect concentration. Band-splitting energy $\Delta E_2$ for $C_2$-type vacancies and the degenerate band energy for $C_3$-type vacancies as functions of (c) superlattice size $N$ and (d) defect concentration. All plots correspond to $3n$ GSLs with $n = 3, \dots, 17$.

### C. The dispersion contour of the cones

The energy dispersion across the BZ can be represented by a Taylor expansion around the Dirac cone to second order, and is given by [1,2,40,41]:

$$E(\mathbf{q}) = \pm v_F|\mathbf{q}| - \left(\pm \frac{3t^2}{8}\sin(3\theta_{\boldsymbol{q}})\right)|\mathbf{q}|^2 \tag{3}$$

where $\mathbf{k} = \mathbf{K} + \Delta\mathbf{q}$ and $|\mathbf{q}| \ll |\mathbf{K}|$ and $\theta_q = \arctan\left(\frac{q_y}{q_x}\right)$. The linear term $v_F|\mathbf{q}|$ gives a circular contour near the cones, while the second term introduces a threefold angular distortion (trigonal warping), making the energies further away from the $E_F$ dominated by the quadratic correction and the contour starts to warp [40,41]. The valley index in the angular term implies that the orientation of the trigonal warping at the K and K′ is reversed [42]. For small momentum near the Dirac cone, this distortion is weak leading to circular contour. The shape of the energy dispersion can be visualized by plotting contours of the discrete $E(\mathbf{k})$. Figs. 7(a-c) show the dispersion contours of the VBM of the GSLs for $3n \pm 1$ and $3n$ sizes, with $n = 3$. Fig. 7(a) shows the dispersion contour of GSLs with $D_{3h(18C)}$ vacancy, where in $3n \pm 1$ the circular contours near the Dirac cones are clearly observed, while the trigonal shape appears at energies further away from Dirac points. Similarly, the dispersion remains circular near the VBM, indicating that it is isotropic near the energy level of the VBM. These patterns are observed in other GSLs with $C_3$ symmetry, as shown in Table II of the Supplemental Material [37], because $C_3$ enforces the $v_F$ to be equal in all directions and prevents anisotropy [43]. In contrast, $C_2$-type vacancies show elliptical contours both at energies close to and far from Dirac cones, indicating anisotropic dispersion due to the breaking of $C_3$ symmetry. In the $3n$ GSLs that preserve $\sigma_d \perp \sigma_v$, such as $D_{2h(16C)}$ in Fig. 7(b) and Table II of the Supplemental Material [37], one elliptical contour at VBM energy is observed, consistent with K and K′ folding to Γ and allowing gap opening. However, in the $C_{2h(4C)}$ case, the mirror constraints are lifted and the cones can move away from Γ, appearing as two elliptical contours near Γ as shown in Fig. 7(c). At higher energies, the anisotropic contribution becomes more pronounced, and there is no symmetry constraint enforcing trigonal warping. This means that the Fermi velocity in GSLs with $C_2$-type vacancies is no longer equal in all directions, suggesting potential for anisotropic transport applications.

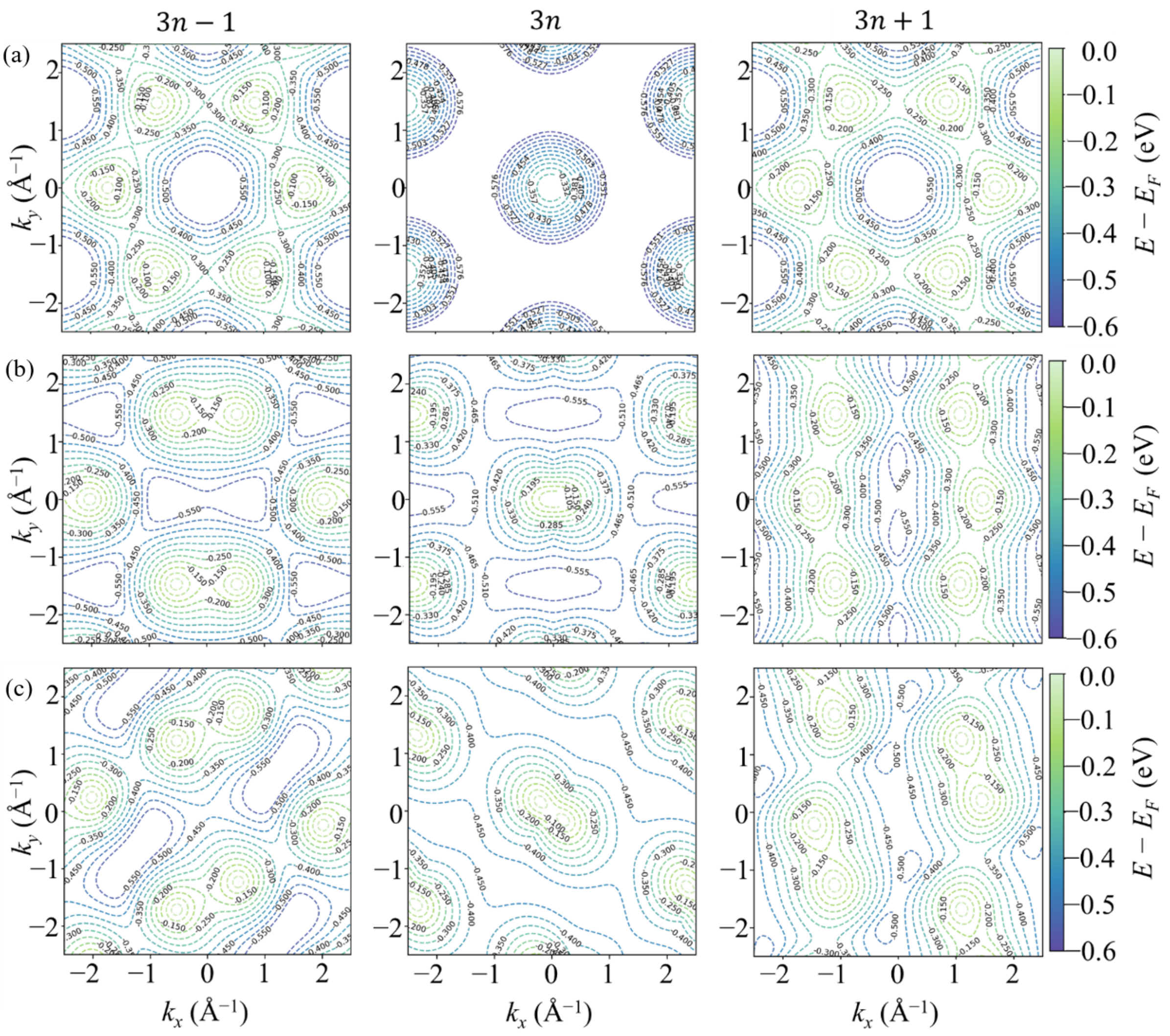

FIG. 7. Energy contour of the VBM in $3n-1$, $3n$, and $3n+1$ GSLs ($n=3$) with (a) $D_{3h(18C)}$, (b) $D_{2h(16C)}$, (c) $C_{2(4C)}$.

### D. Effect of periodicity-perturbing vacancy displacements and missing vacancies on the band gap

In experiments, achieving perfect long-range periodicity of π-vacancies in graphene is challenging, and random positional disorder is expected in practical patterning approaches. In addition, vacancies may be missing during the fabrication process. To assess the robustness of the band gap opening against these imperfections, disorder simulations for the $D_{6h(6C)}$ vacancy in the 9×9 GSLsuperlattice. The perfect GSL has a defect concentration of 3.7% and a band gap of 314 meV. We construct an enlarged GSL from the $3n$ GSL with $n = 3$, by repeating it four times along each lattice direction, resulting a GSL with 16 $D_{6h(6C)}$ vacancies as shown in Fig. 8. We displace or remove up to eight out of the 16 individual defects, as shown in Figs. 8(b,c) and Fig. 8(d), respectively, thereby perturbing the defect translational symmetry. Vacancy displacements were introduced by shifting the entire vacancy by one or two primitive lattice vectors of graphene from the perfect location, with the direction randomly selected from $\pm\mathbf{a_1}, \pm\mathbf{a_2}$, and $\pm(\mathbf{a_2} - \mathbf{a_1})$, as shown in Figs. 8(b) and 8(c), respectively. For each degree of disorder, five independent random configurations in terms of defect and its displacement direction were considered, as summarized in Table IV of the Supplemental Material [37]).

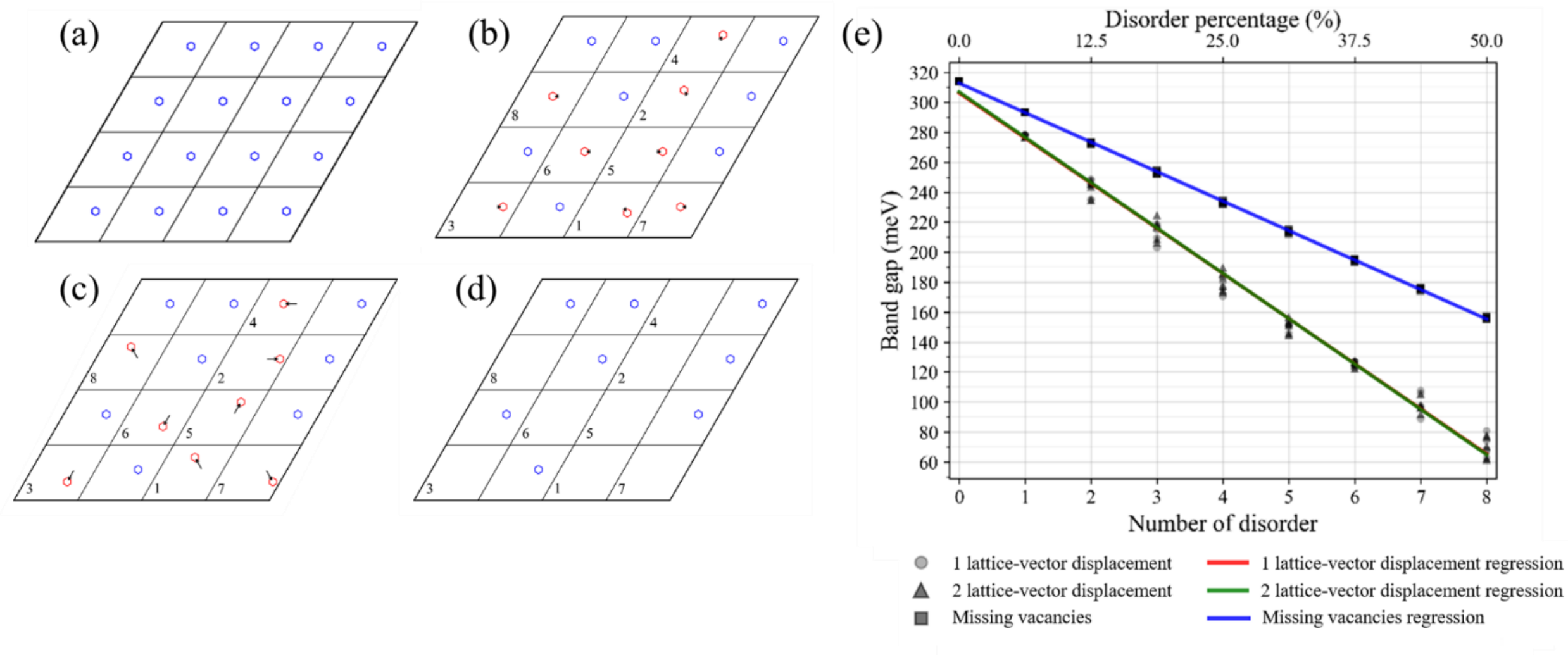

FIG. 8. (a) Ordered enlarged GSL containing 16 $D_{6h(6C)}$ vacancies, with each square representing one original GSL region containing one vacancy. (b) Representative displacement by one lattice vector from configuration 1. (c) Representative displacement by two lattice vectors from configuration 2. Red hexagons and arrows indicate the positions and directions of displaced vacancies. (d) Representative missing vacancy from configuration 3, with a blank box indicating a missing vacancy. (e) Band-gap reduction as a function of the number of displaced or missing

vacancies for different random configurations, along with the regression lines to show the band gap suppression trends.

Fig. 8 (e) shows that the band gap decreases linearly as a function of the number of displaced or removed $D_{6h(6C)}$ vacancies. For displaced defects, the band gap decreases by 30 meV per defect and nearly closes at a 50% defect density. Missing vacancies also reduce the band gap linearly with concentration, but by a smaller amount of 20 meV per defect (as shown in Fig. S3 and Table IV in the Supplemental Material [37]).

### E. Symmetry-based guidelines for band gap opening in the GSL with periodic $\pi$-vacancy

A general symmetry-based guideline for band-gap opening in a GSL with periodic $\pi$-vacancies can be formulated from our results, as shown in Fig. 9. At the TB level considered here, two base requirements ensuring that the $\pi$-network modulation can open the band gap in the GSLs are equal removal of sites of A and B sublattices, which avoids zero modes at $E_F$, and a uniform on-site energy on all remaining sites ($\varepsilon_i = \varepsilon_0$) to keep the sublattice symmetry. Under these conditions, the gap formation is controlled mainly by the symmetry constraints due to the $\pi$-vacancy modulation. The translational symmetry enforces GSL folding, and a band gap can be obtained effectively when the GSL size satisfies $3n$. The perturbation can occur when both K and K$'$ valleys are folded to the TRIM point of $\Gamma$. In contrast, for $3n \pm 1$, the two valleys remain separated in momentum space, leaving the Dirac cone survive in the momentum space. Therefore, in practice, $3n$ GSL size is needed for $\pi$-vacancy gap engineering.

In these GSLs sizes, the decisive factor becomes whether the vacancy pattern pins the folded cones at Γ and allows the degeneracy to be lifted. Our results indicate two practical symmetry routes. The most robust route is to design a vacancy motif that preserves $C_3$ symmetry. In this case, threefold rotation constrains the cone positions remain locked to high-symmetry points in the mBZ, enabling the gap opening at Γ in $3n$ GSLs and can be captured directly from high-symmetry sampling. Consistent with this, the $C_3$-type vacancies studied in this report generate larger band gaps at comparable defect concentrations than $C_2$-type vacancies. In particular, $D_{6h(6C)}$ is the most efficient motif in this study, achieving a large gap at a relatively low defect concentration, illustrating a benefit of preserving $C_3$. The second symmetry route is available for vacancy motifs without $C_3$. In these cases, a gap can still open in $3n$ GSLs if two perpendicular mirror planes $\sigma_d \perp \sigma_v$ are preserved in the mBZ. The remaining mirror planes restrict the Dirac cone shift along the $k_\perp$, as consequences the cones are located at Γ in $3n$ GSLs, enabling gap opening. The perpendicular mirror planes $\sigma_d \perp \sigma_v$ are a sufficient condition, but not necessary. However, in comparison with the $C_3$-preserving cases, the resulting gaps are smaller due to the nondegenerate bands. Hence, for experimental realizations where perfect periodicity is difficult to achieve, introducing a $C_3$-type vacancy is not only a practical route for band gap opening, but also a route to a robust band gap against displacement. The $\pi$-vacancies are models can represent various defect types, such as hydrogenation leading to $sp^3$ functionalization [12,13,15] or BN substitution [5,6,7]. These changes maintain the graphene bonding framework but can result in local deformation, which in turn can quantitatively change the electronic structure, but not the overall character of the band structure. In contrast, atomic vacancies give rise to strong relaxation effects with severe impact on the electronic structure of the GSL as discussed elsewhere[44,45].

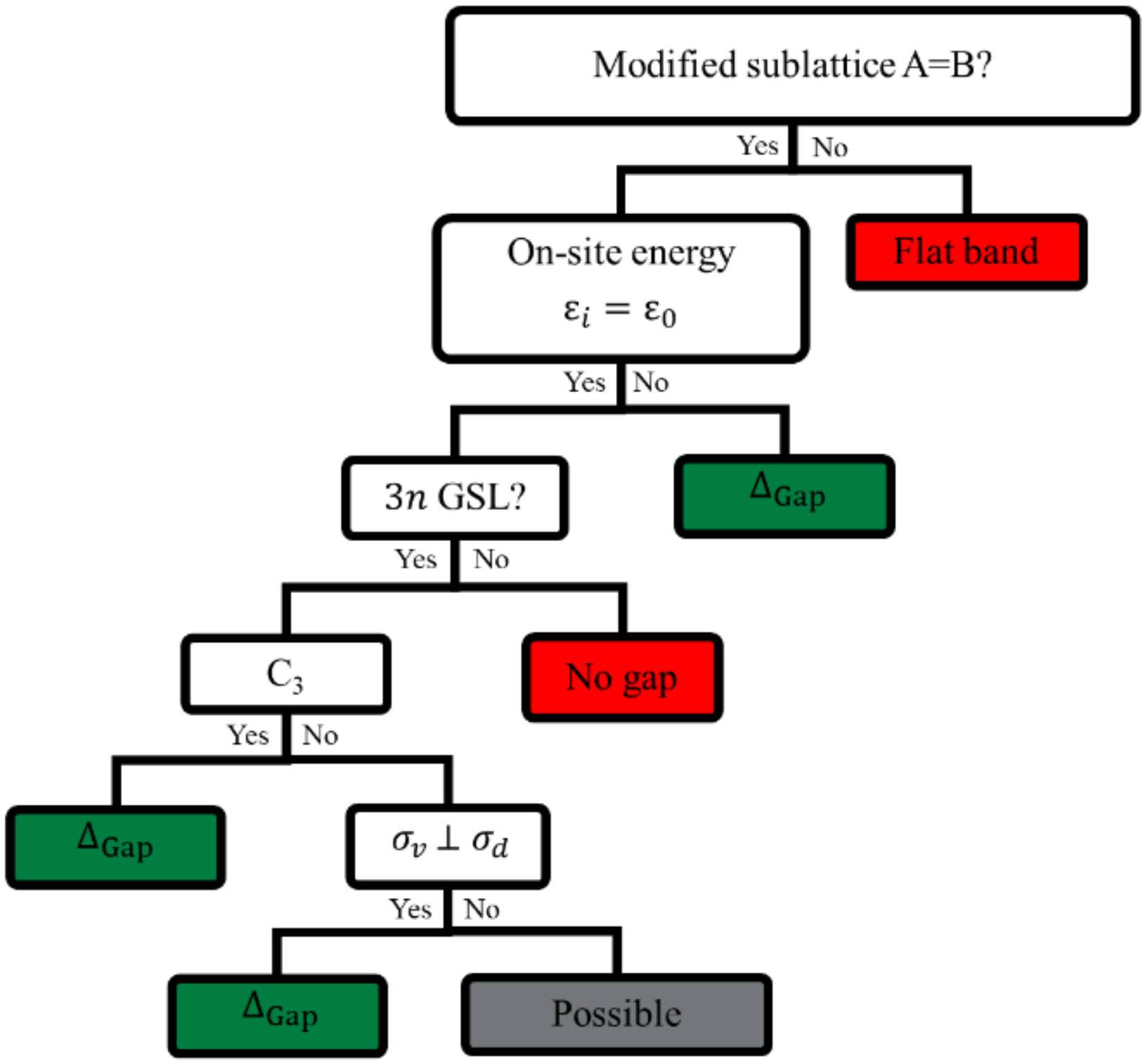

FIG. 9. An mBZ symmetry-based guideline to open the band gap in the GSL with periodic π-vacancy.

## IV. CONCLUSIONS

In summary, we showed that periodic π-vacancies in GSLs can induce band gap opening. Importantly, these results can be applied in broader defect engineering strategies, such as substitution, functionalization, or missing carbon sites. We found the following rules for opening a band gap in GSLs with periodic π-vacancies: equal removal of sublattice A and B sites, with a uniform on-site energy at remaining sites, and a superlattice size of $3n$ is needed for folding the K and K′ to Γ. A feasible way to open the band gap is to preserve $C_3$ symmetry, which globally stabilizes the Dirac cone position at high-symmetry points, allowing K and K′ to be pinned at Γ in $3n$ GSLs. $C_2$-type vacancies can open a band gap if the perpendicular mirror planes, $\sigma_v \perp \sigma_d$, are preserved in momentum space as a sufficient (but not necessary) condition. The $C_3$-type vacancies yield larger band gap, with $D_{6h(6C)}$ as the most efficient motif in terms of defect concentration, indicating the practical advantage of preserving $C_3$ symmetry. Vacancy disorder involving displacements by one or two primitive lattice vectors of graphene or missing $D_{6h(6C)}$ vacancies break the $C_3$ symmetry, lift the degeneracy, and reduce the gap toward the pristine graphene limit. For experimental realizations where perfect periodicity is difficult to achieve, introducing a $C_3$-type vacancy motif is therefore a practical route to large band gaps while maintaining robustness to displacement. We would like to point out that bottom-up precision synthesis approaches to functionalized graphene are a promising route to maintain the structural stability and a nearly linear dispersion of the electronic structure while opening a sizeable band gap of more than 500 meV in graphene, which is sufficient for stable electronics at room temperature.

## ACKNOWLEDGMENTS

This work was funded by the Deutsche Forschungsgemeinschaft (DFG, German Research Foundation) within projects RTG 2861 and HE 3541-47-1. The authors gratefully acknowledge the computing time provided to them on the high-performance computer at the ZIH.

## DATA AVAILABILITY

There are no publicly available research data supporting this manuscript. Requests for further data should be sent to the authors.

# Supplemental Material

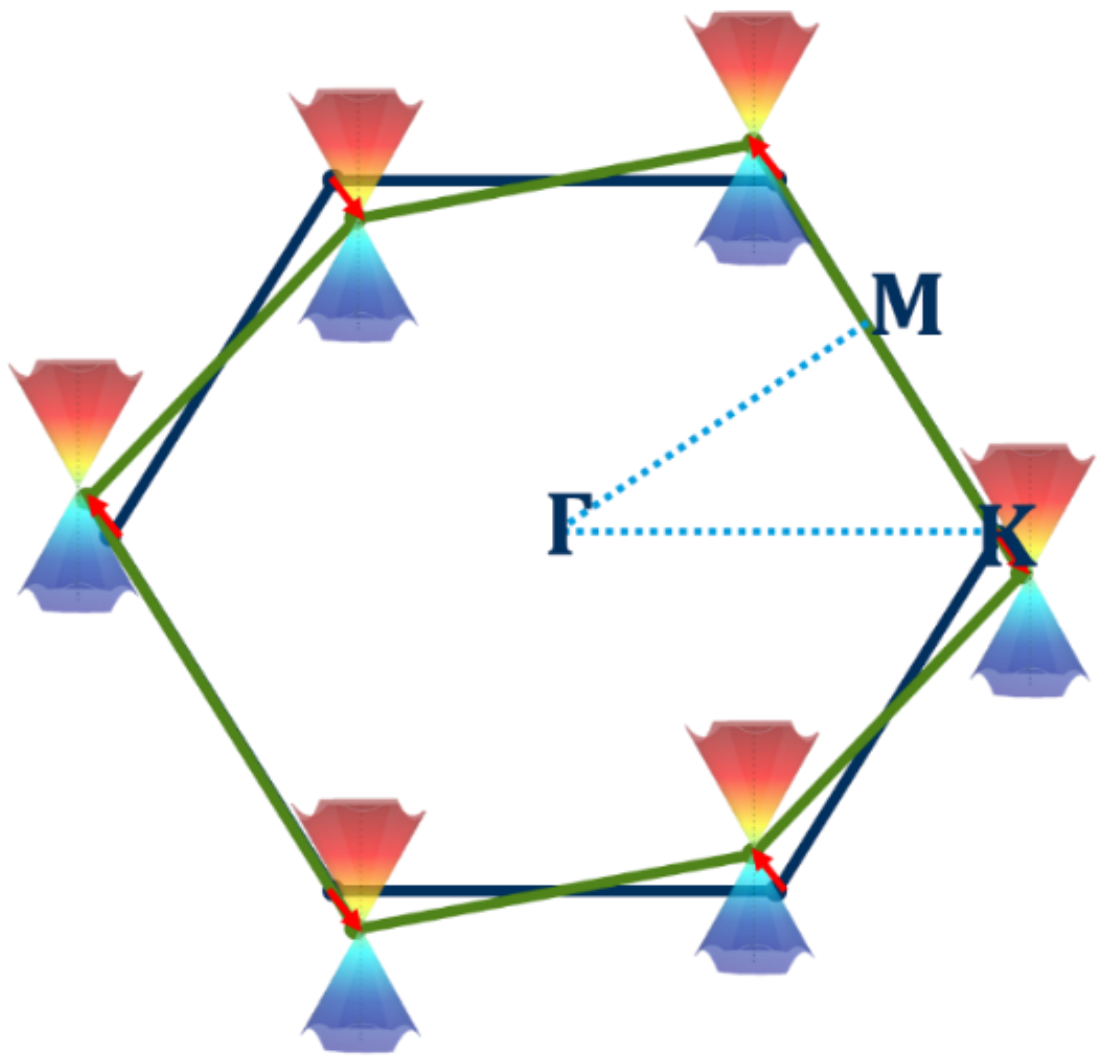

FIG. S1. Dirac cone shifting in $D_{2h(2C)}$ 8x8 GSL

Table I. The 1D and 2D band structure plots of GSL with periodic vacancy for $n = 3$, $E_F$ is denoted by the red dash.

| | |
|---|---|
| 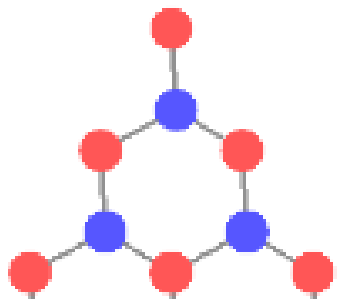 $D_{3h(16C), A \neq B}$ | 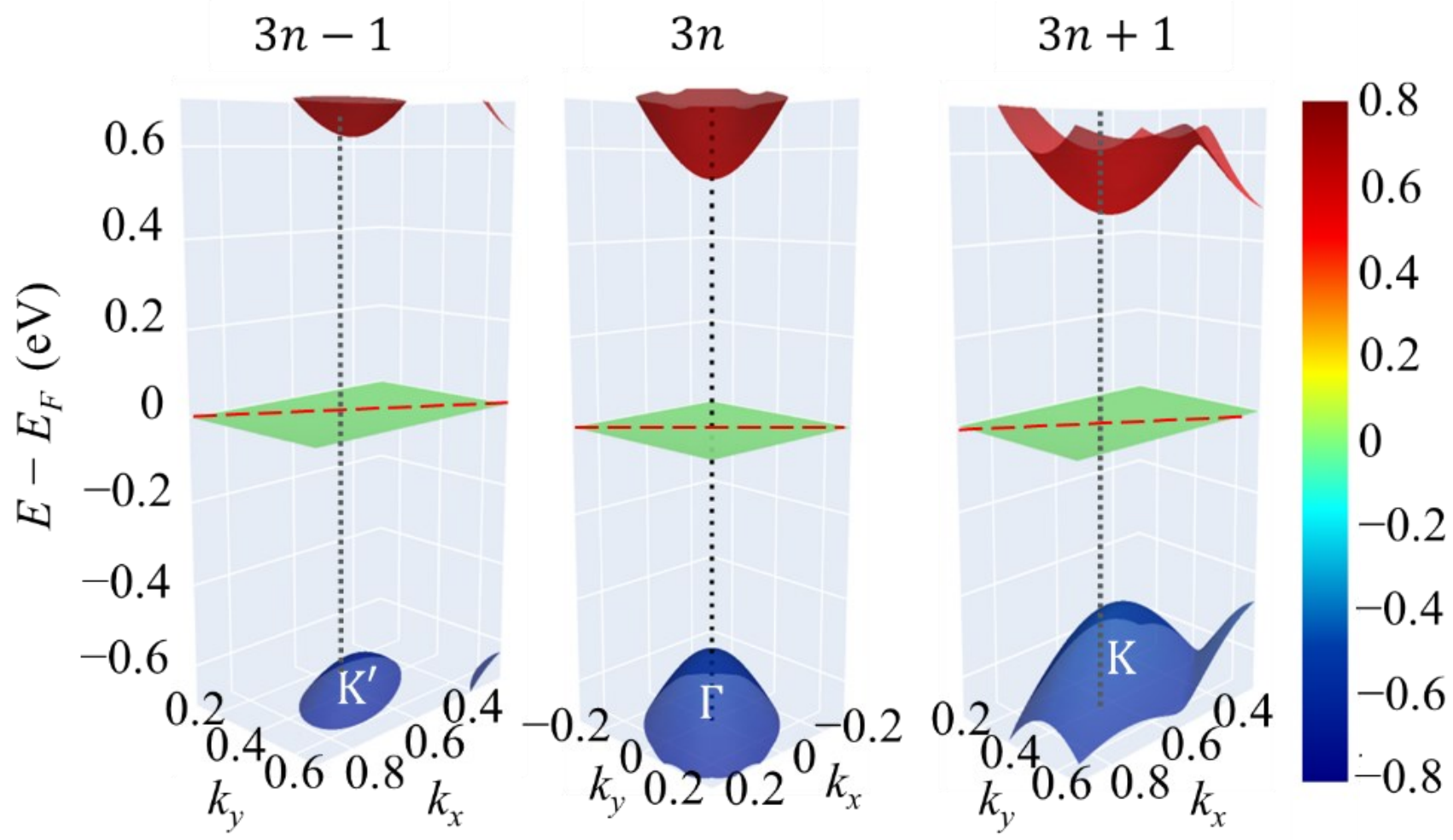 |
| | 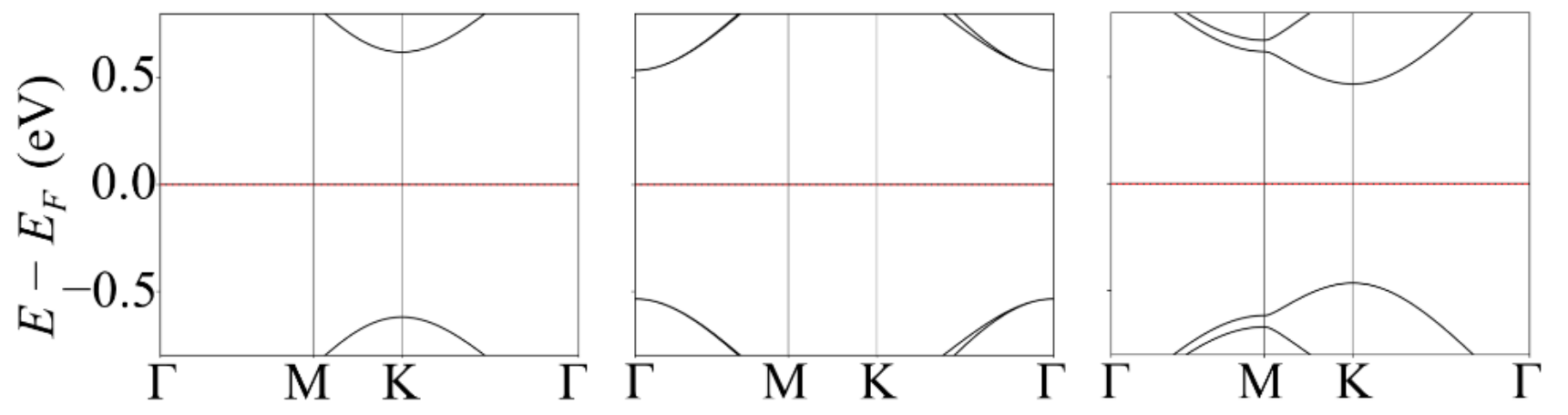  |
| 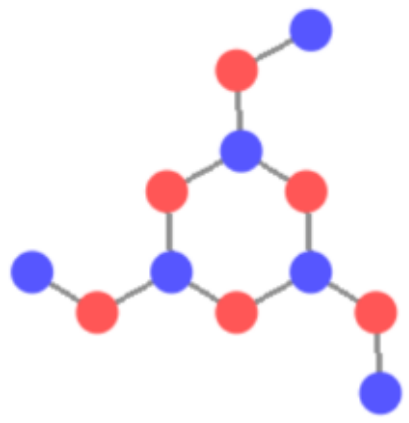<br>$C_{3(12C)}$ | 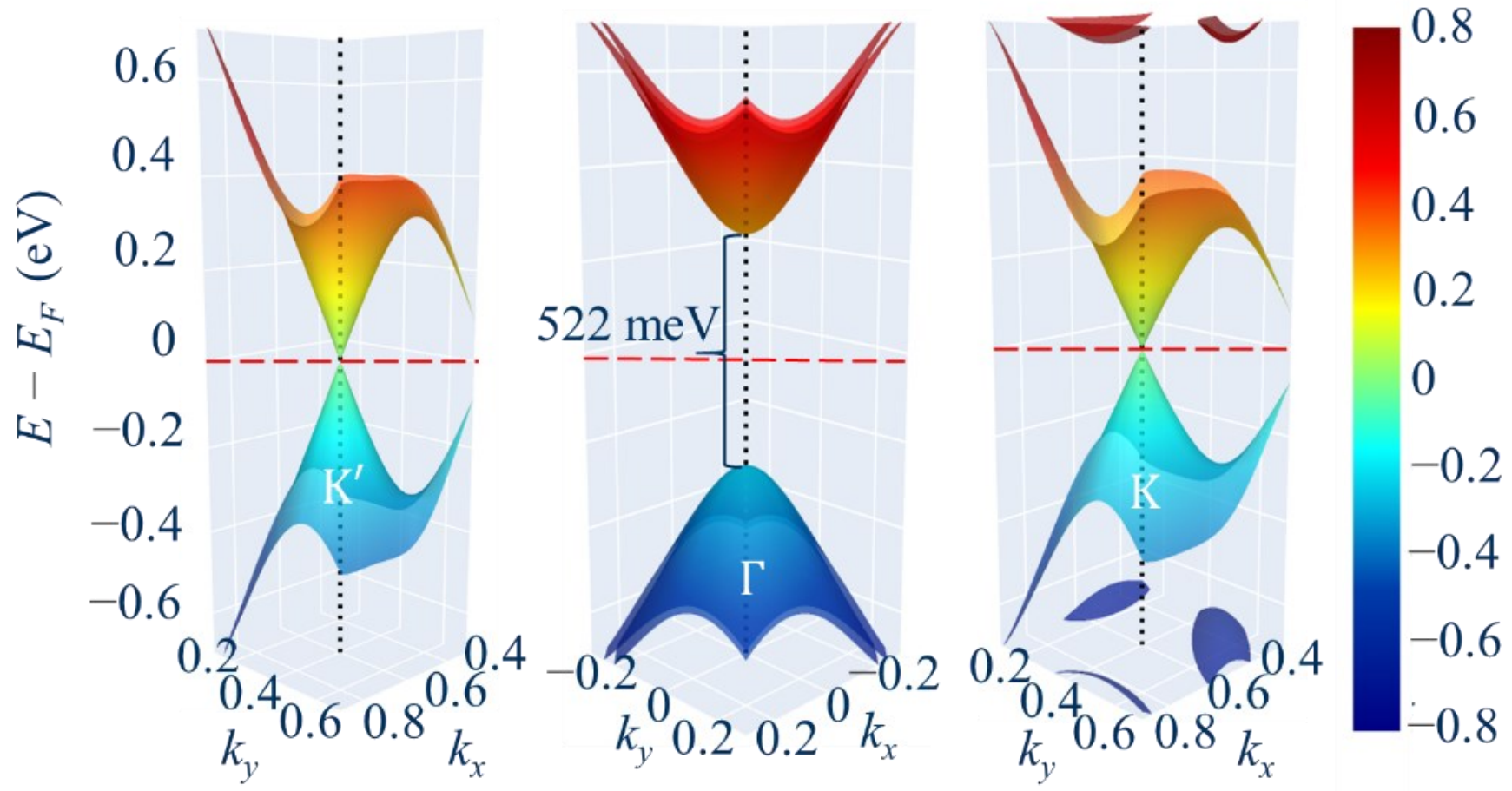 <br>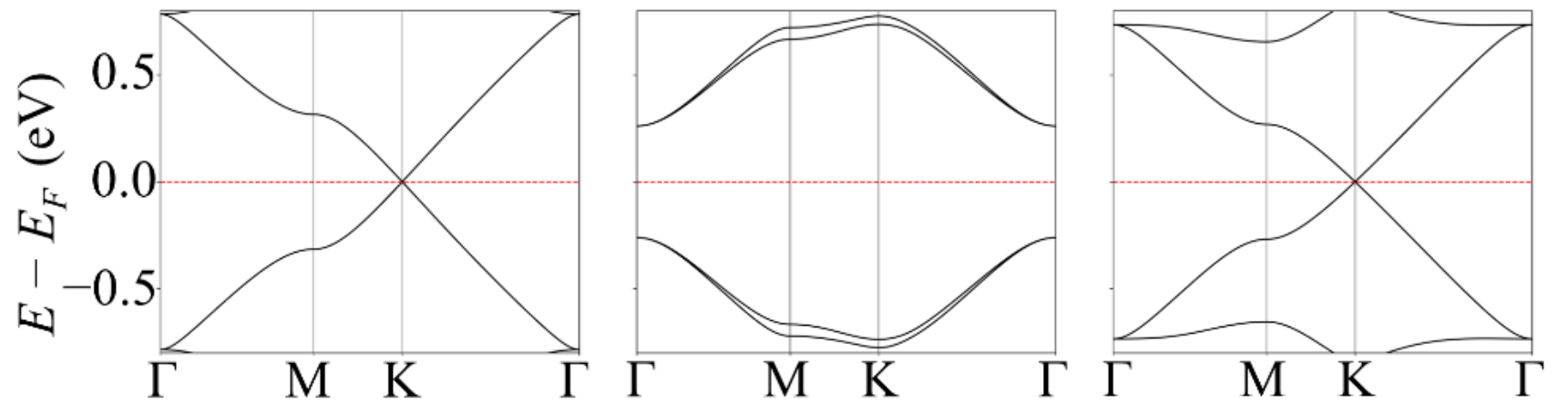  |
| 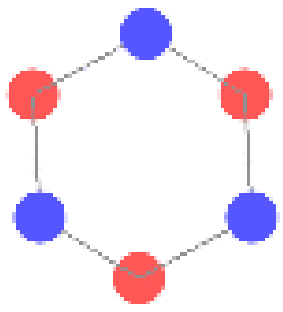<br>$D_{6h(6C)}$ | 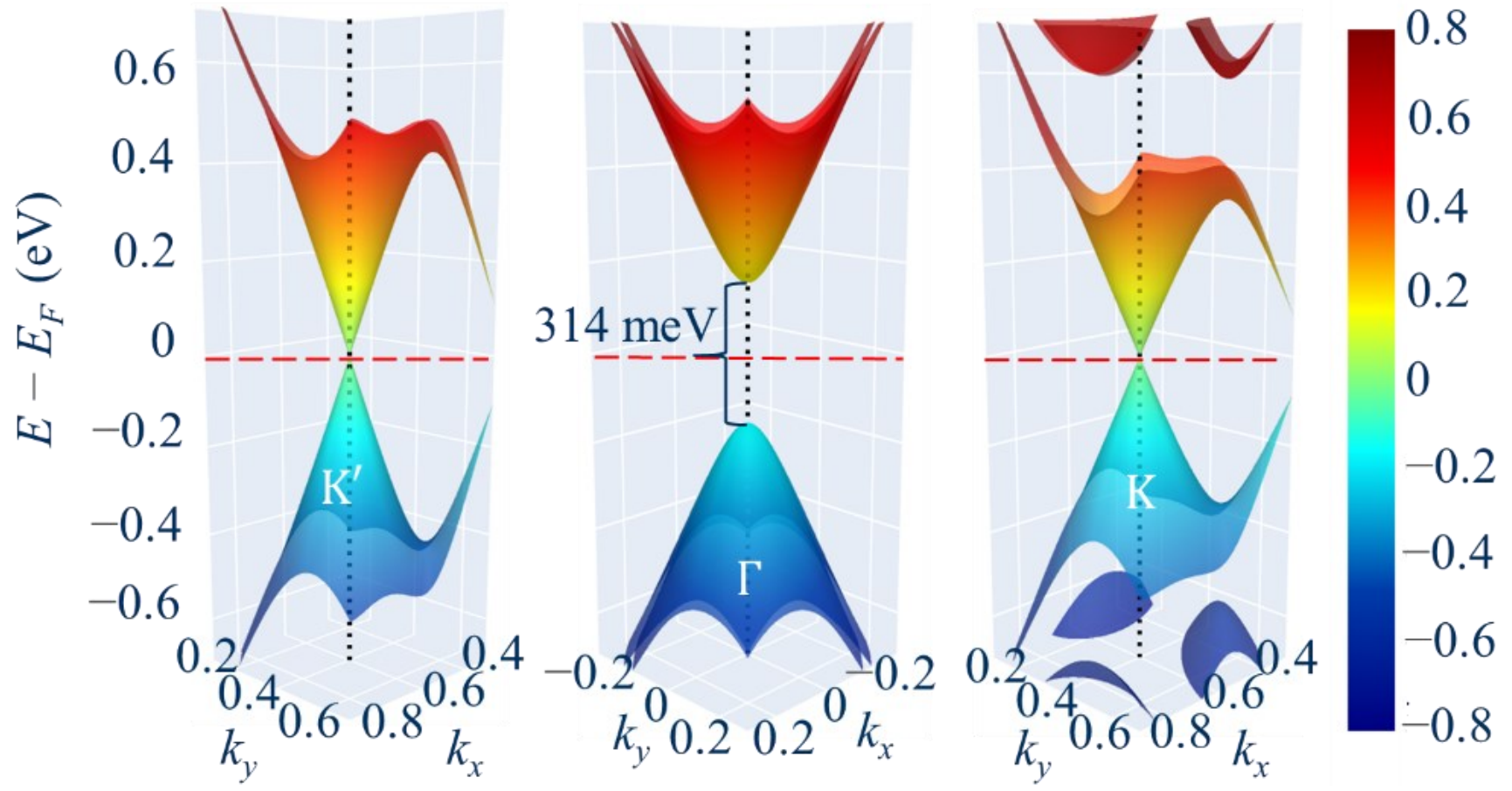  |

<table>
<tr>
<td></td>
<td>
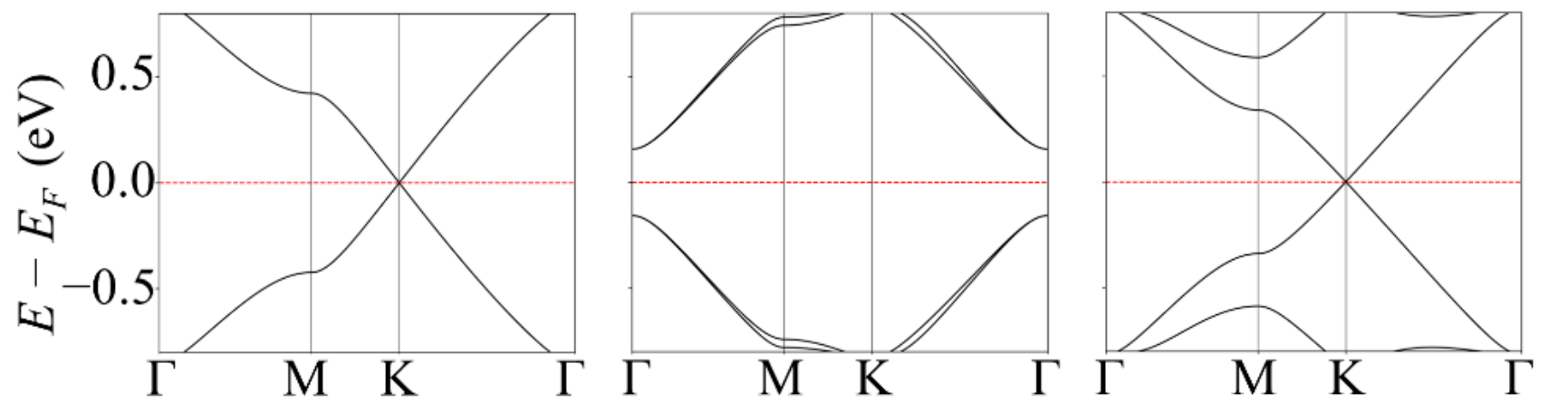

</td>
</tr>
<tr>
<td>

D2h(2C)
</td>
<td>
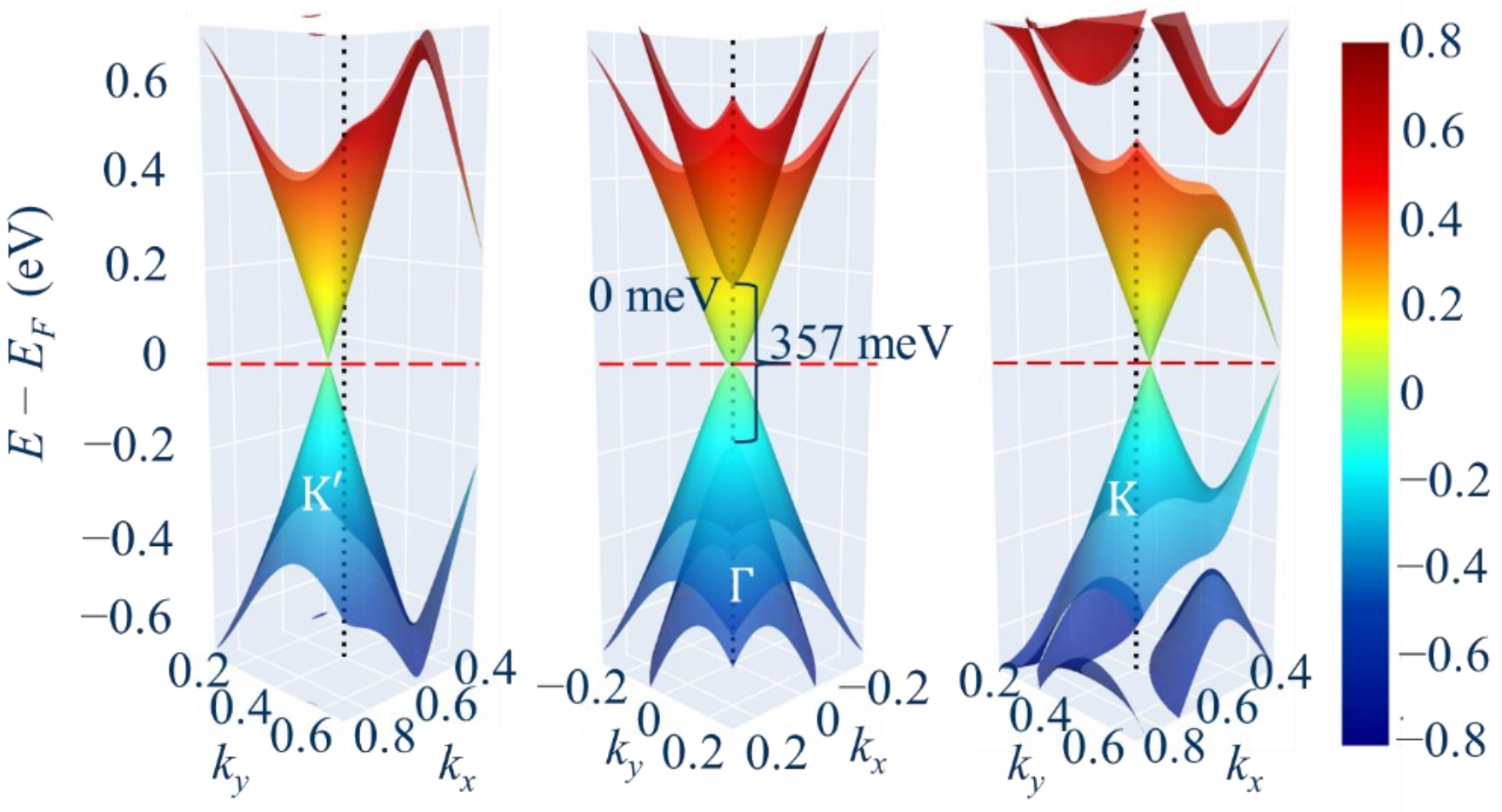

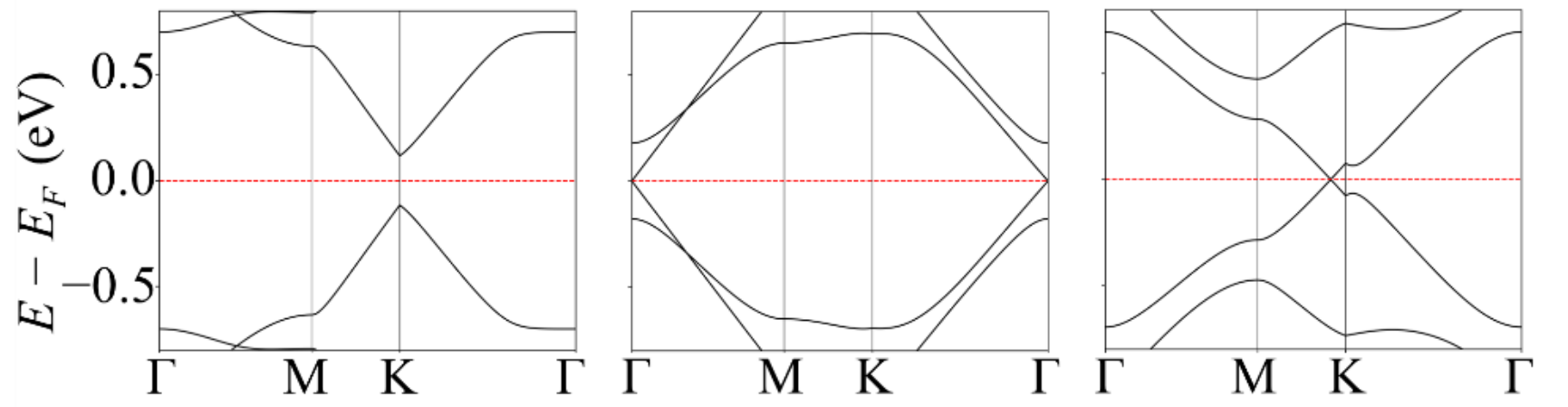

</td>
</tr>
<tr>
<td>
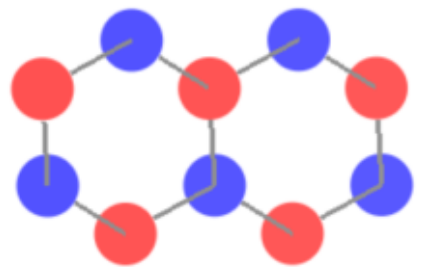

D2h(10C)
</td>
<td>
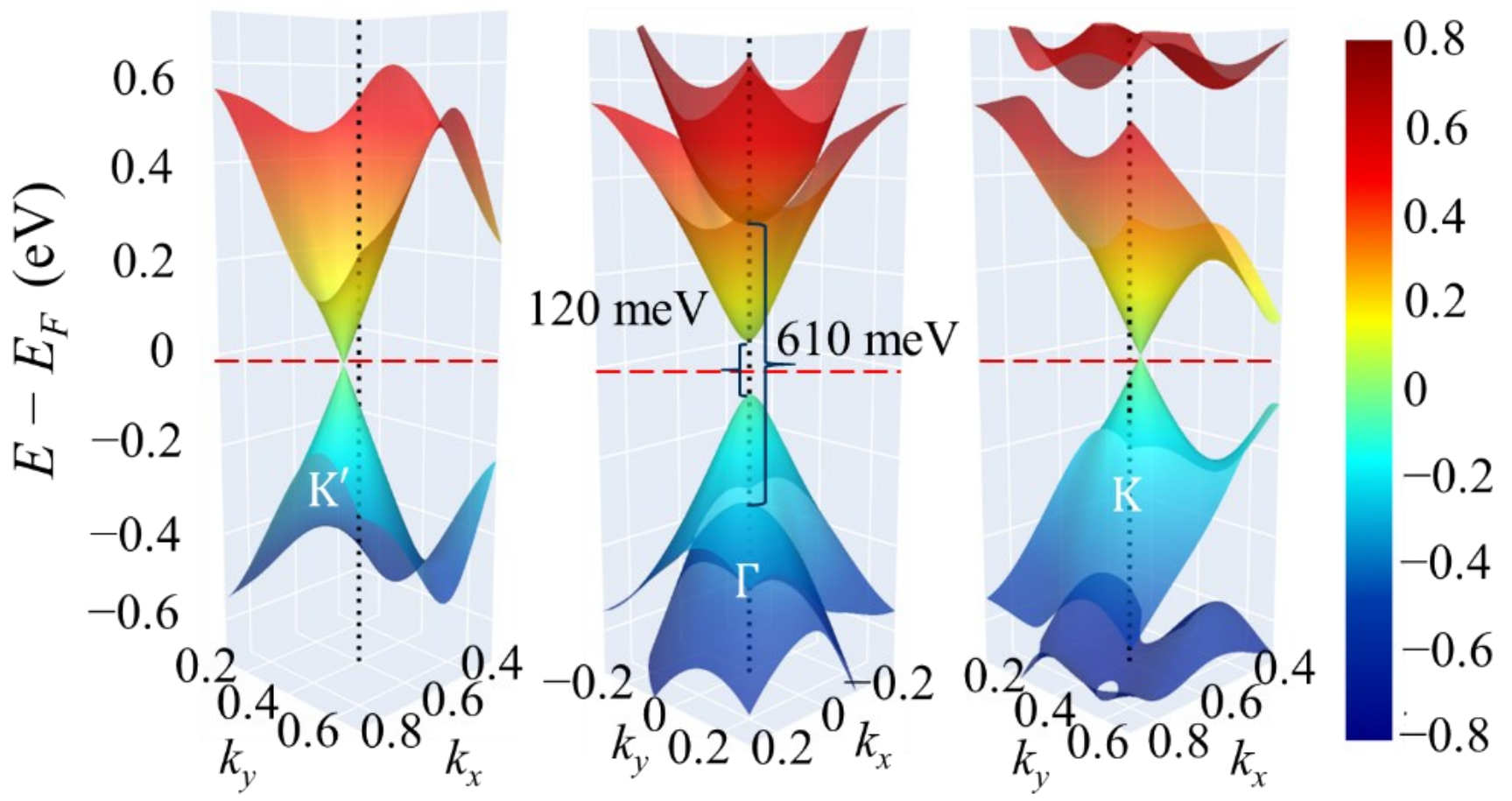

</td>
</tr>
</table>

| | |
|---|---|
| | 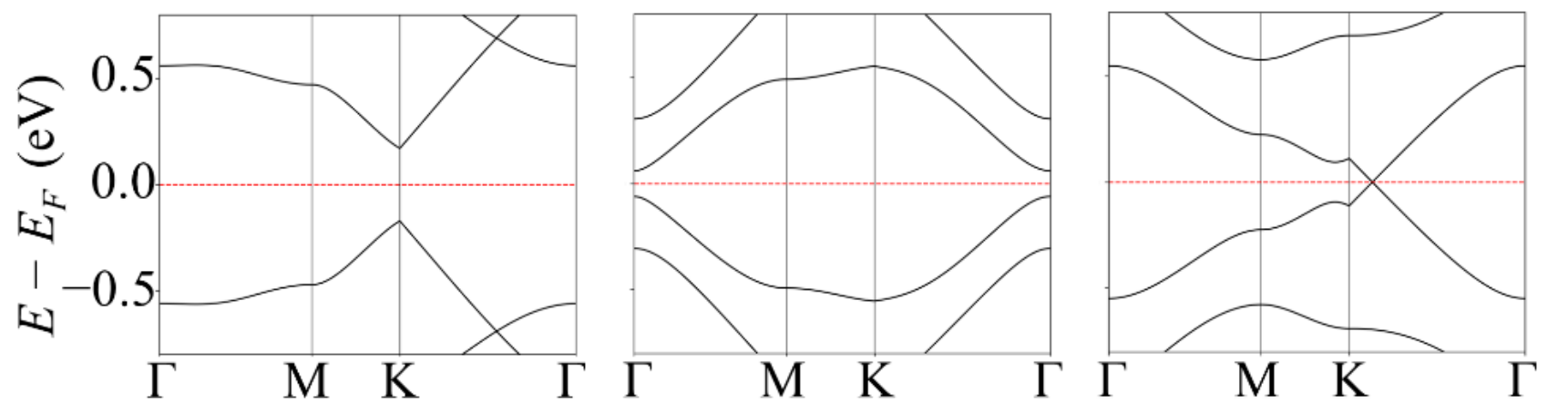  |

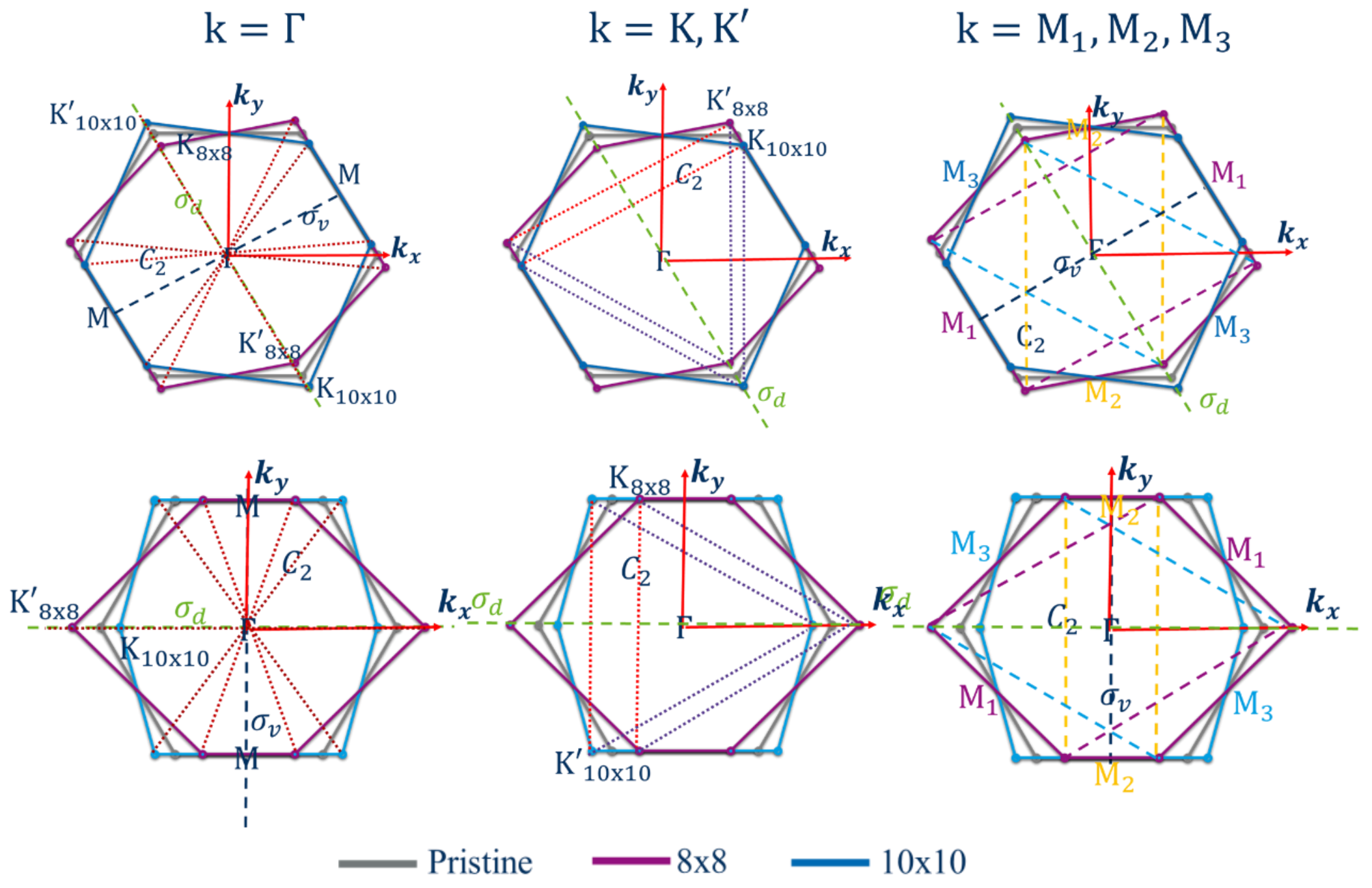

FIG. S2 Reduced little groups at the high-symmetry points ($\Gamma$, K, K$'$, $M_1$, $M_2$, and $M_3$) in the GSL with (a) $D_{2h(2C)}$ and (b) $D_{2h(10C)}$ vacancies for superlattice sizes 8×8 and 10×10.

Table II. Energy contour of VBM of GSL with periodic vacancy

| | |
|---|---|
| Pristine graphene | 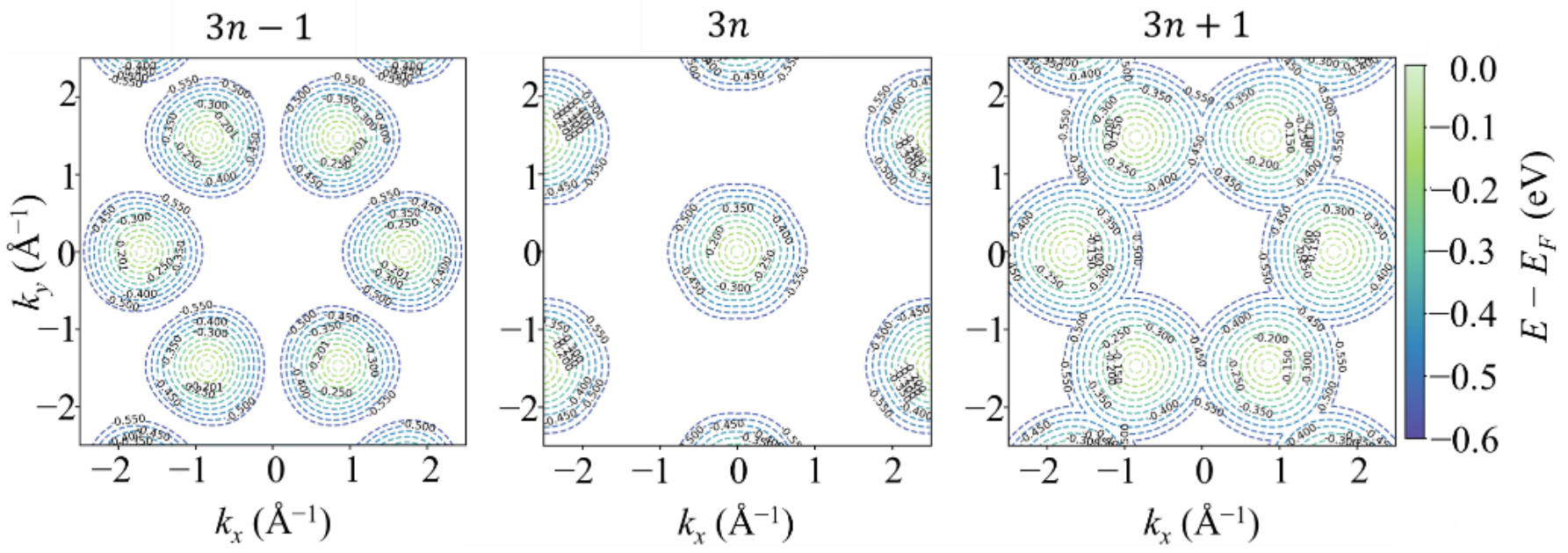  |
| 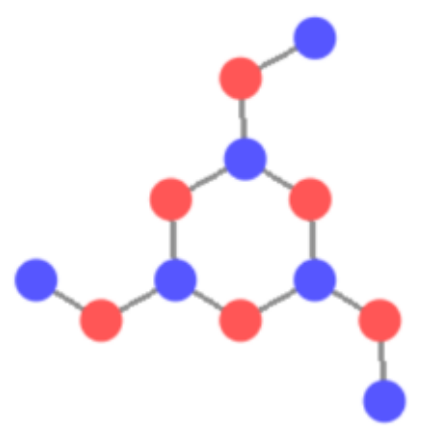 C$_{3(12C)}$ | 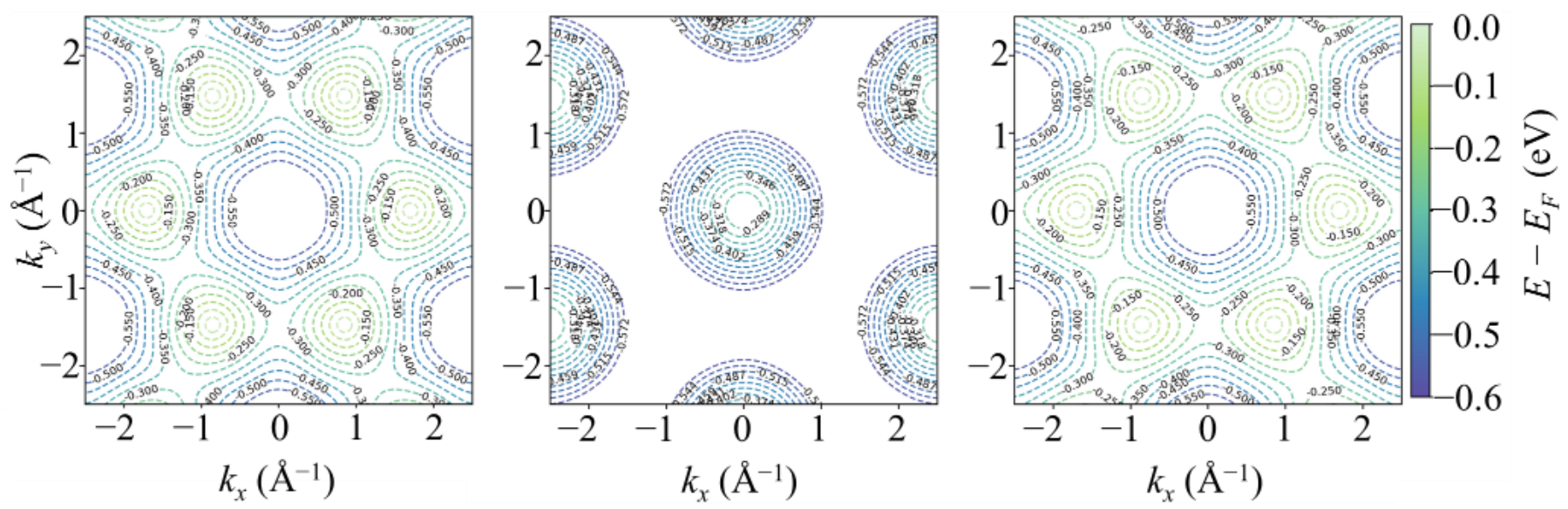  |
| 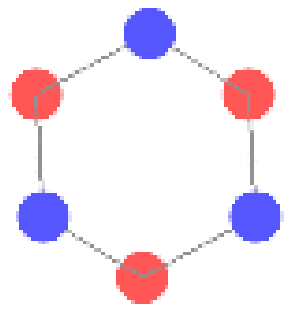 D$_{6h(6C)}$ | 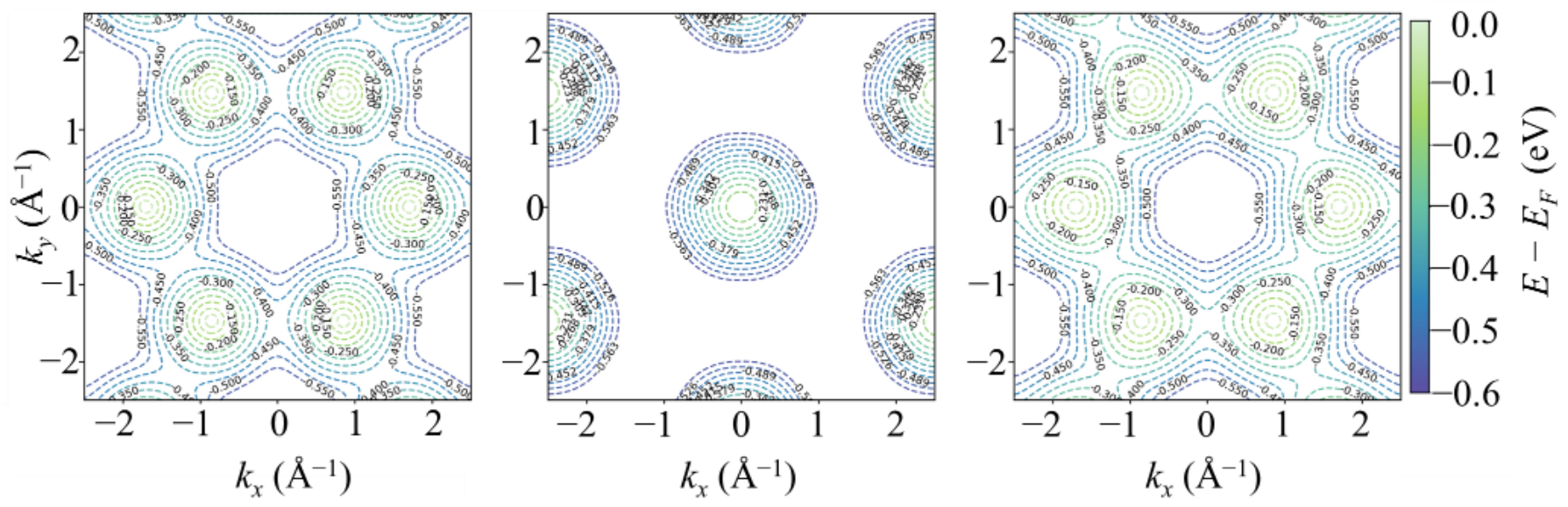  |
| 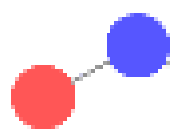 D$_{2h(2C)}$ | 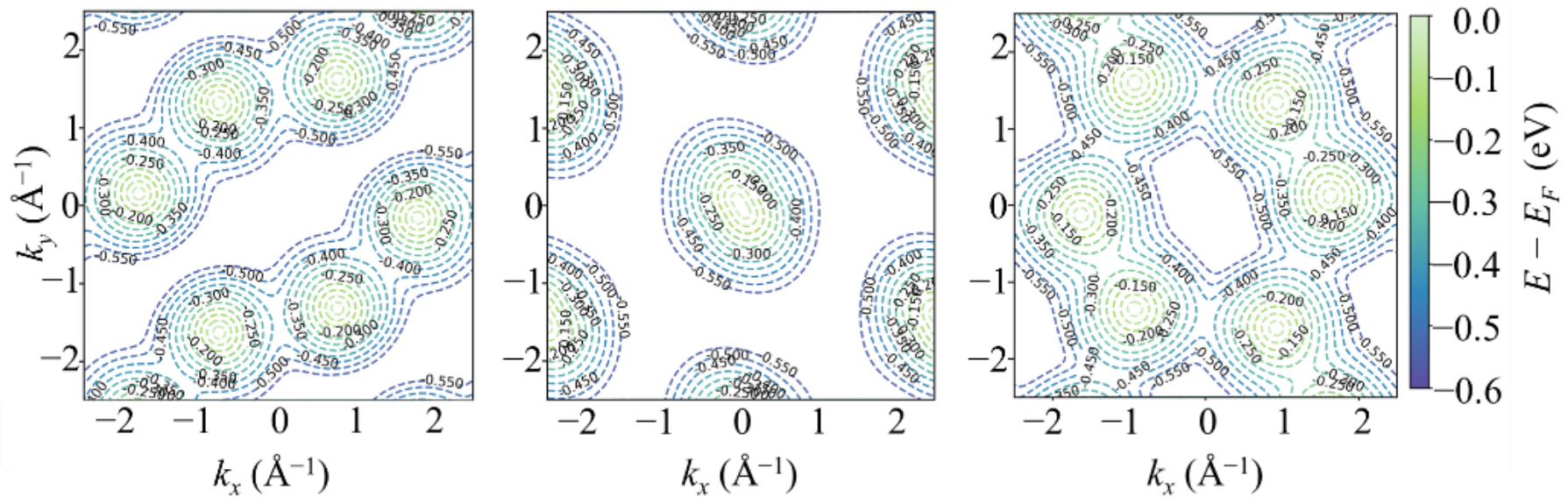  |
| 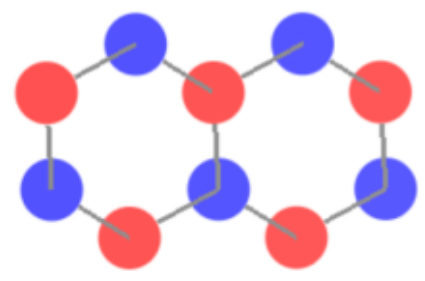 D$_{2h(10C)}$ | 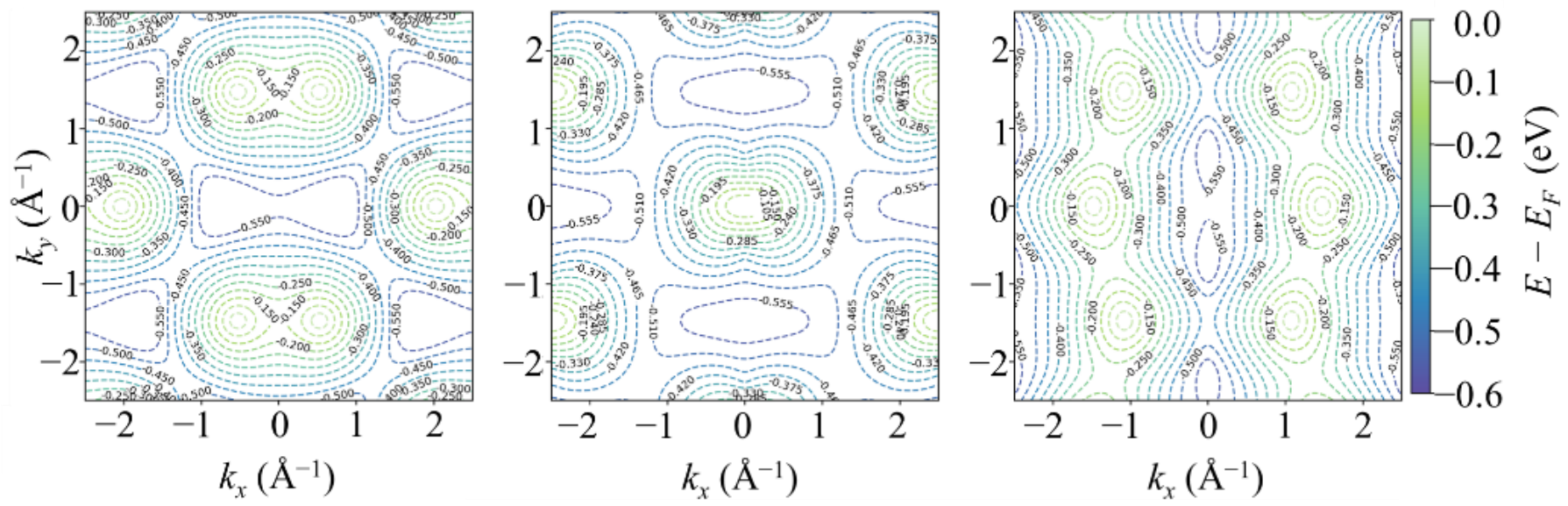  |

Table III. Linear fitting of $\Delta E_1$ as function of defect concentration for each π-vacancy motif

| **Vacancy** | **Linear Fitting of $\Delta E_1$(meV)** | **$R^2$** |
|---|---|---|
| D$_{6h(6C)}$ | $y = 84.832x + 0.979$ | 1.000 |
| $C_{3h(12C)}$ | $y = 70.388\,x + 2.028$ | 1.000 |
| $D_{3h(18C)}$ | $y = 55.179\,x + 0.162$ | 1.000 |
| $D_{(2h(2C))}$ | - | - |
| $D_{(2h(10C))}$ | $y = 19.538\,x - 0.221$ | 1.000 |
| $D_{(2h(16C))}$ | $y = 11.963\,x - 0.096$ | 1.000 |

Table IV. Disorder configurations for vacancies displaced by one or two primitive lattice vectors of graphene, with the direction randomly selected from $\pm\mathbf{a_1}, \pm\mathbf{a_2}$, and $\pm(\mathbf{a_2} - \mathbf{a_1})$, and for missing vacancies

| Displacement by *a* | Dsiplacement by 2*a* | Missing vacancies |
|---|---|---|
| Configuration 1<br>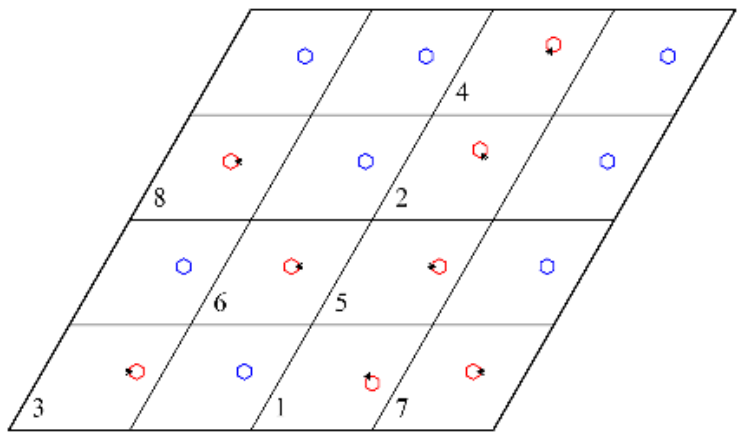 | 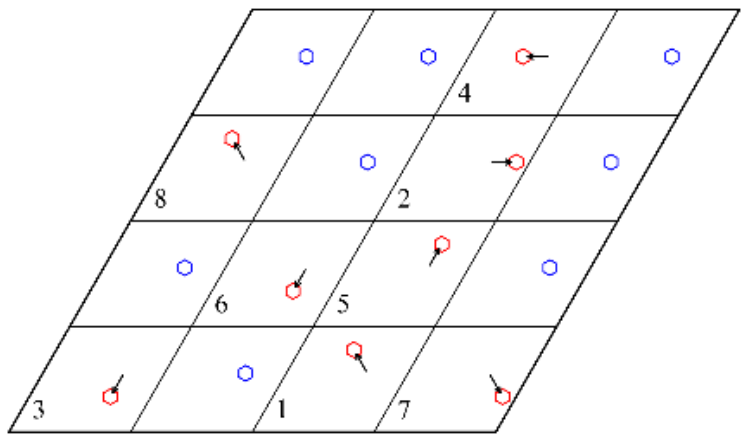 | 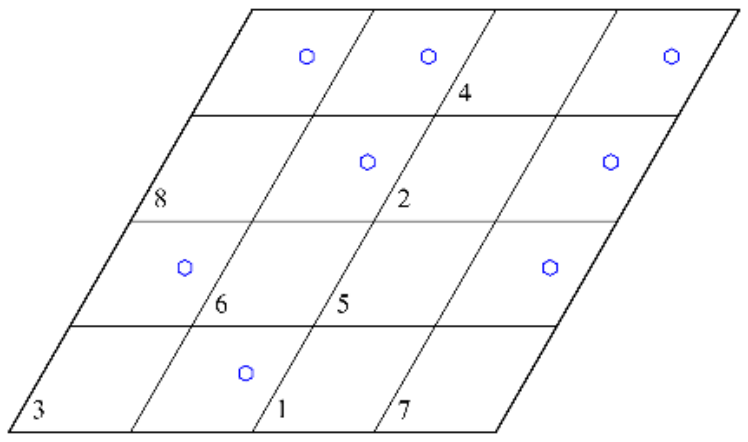 |
| Configuration 2<br>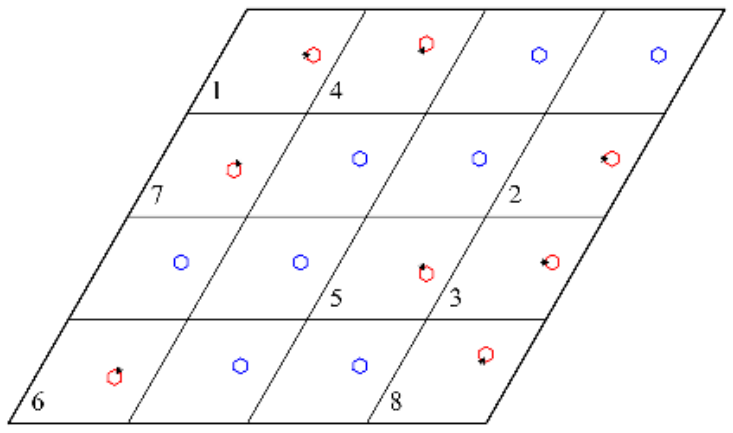 | 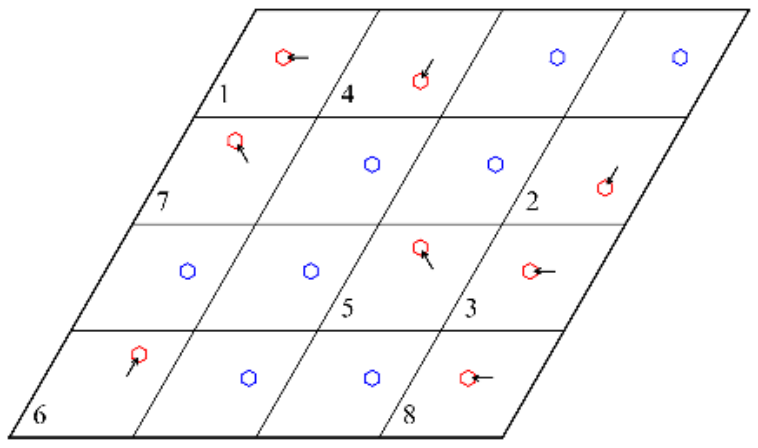 | 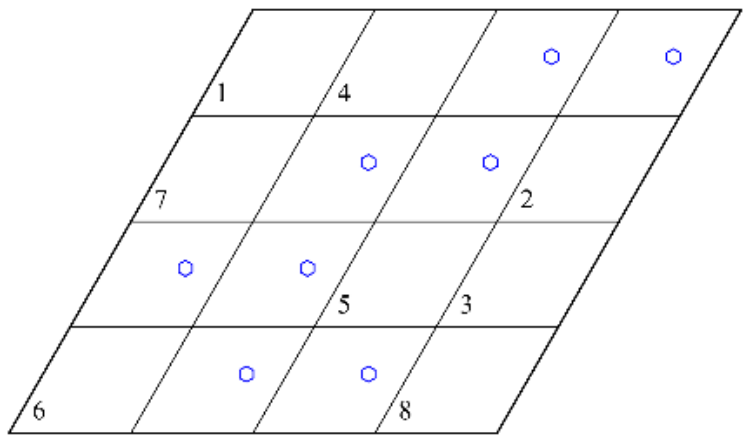 |
| Configuration 3 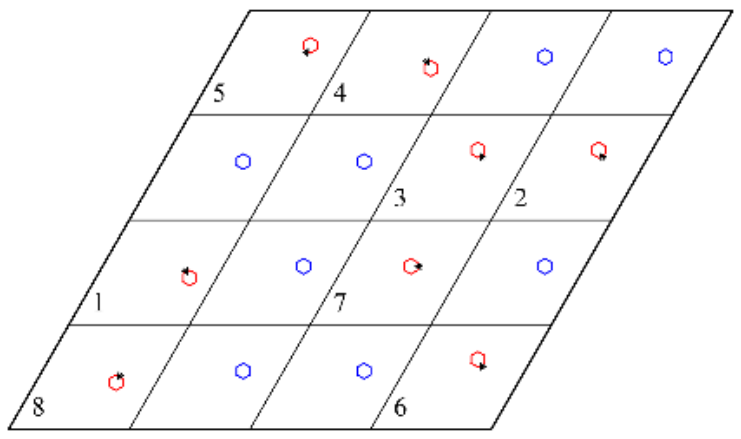 | 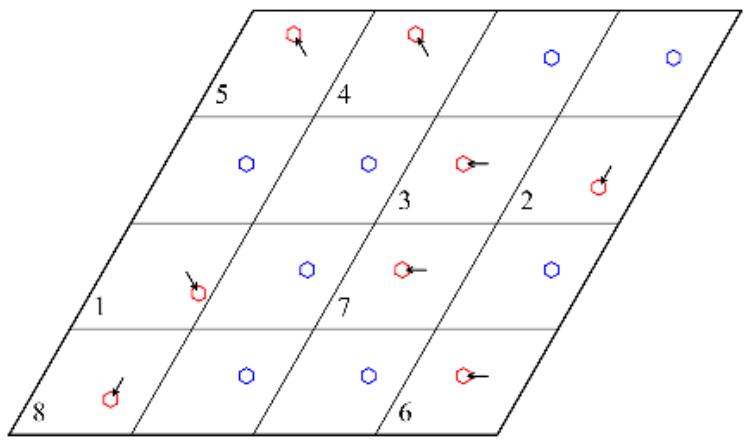 | 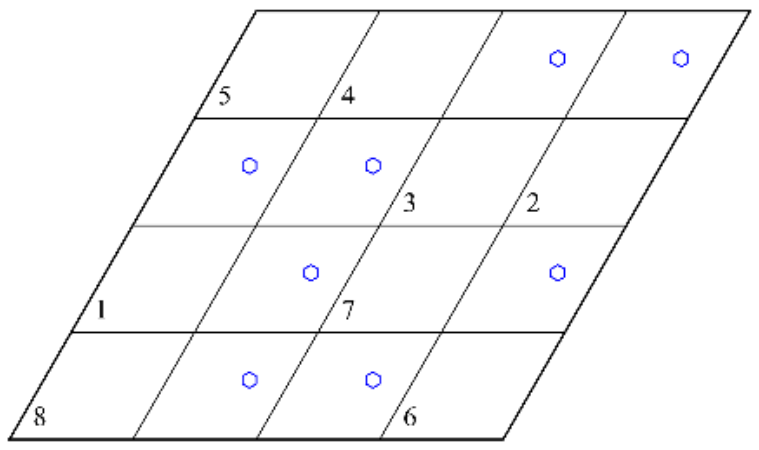 |
| Configuration 4 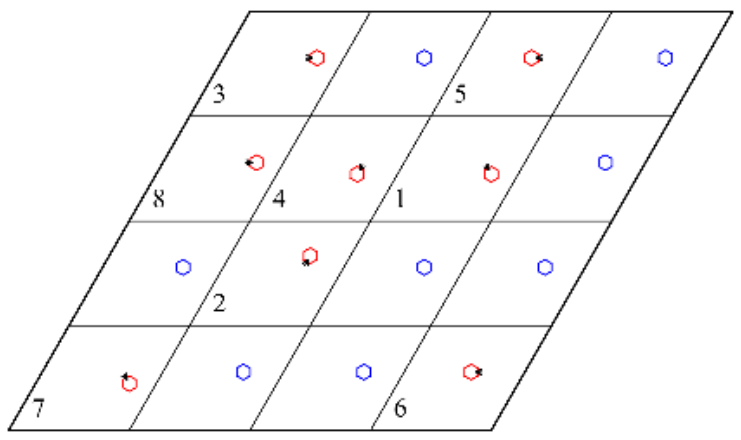 | 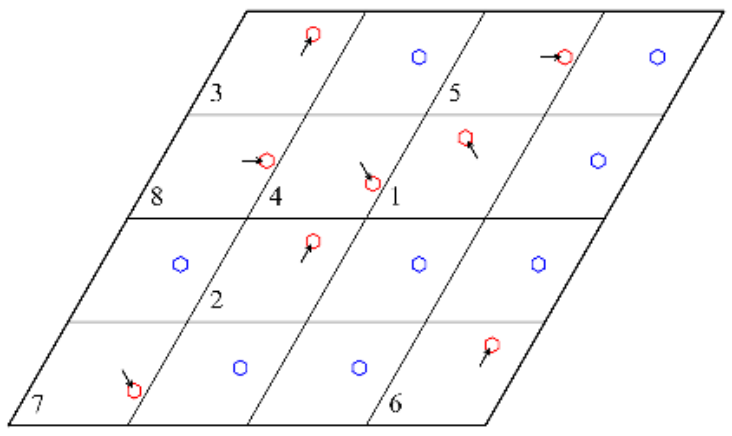 | 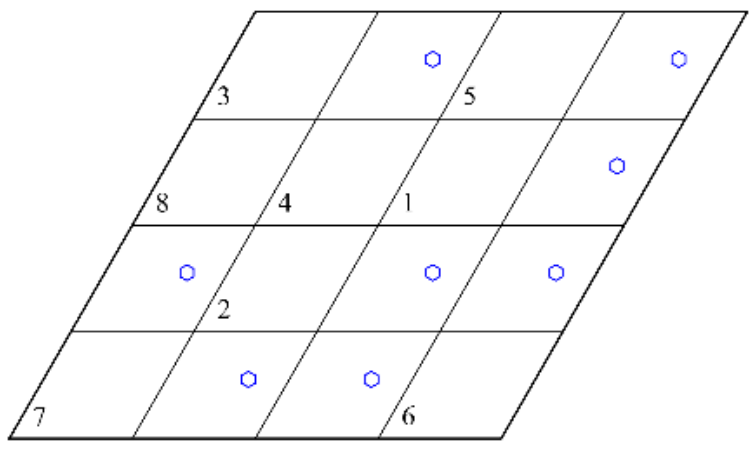 |
| Configuration 5 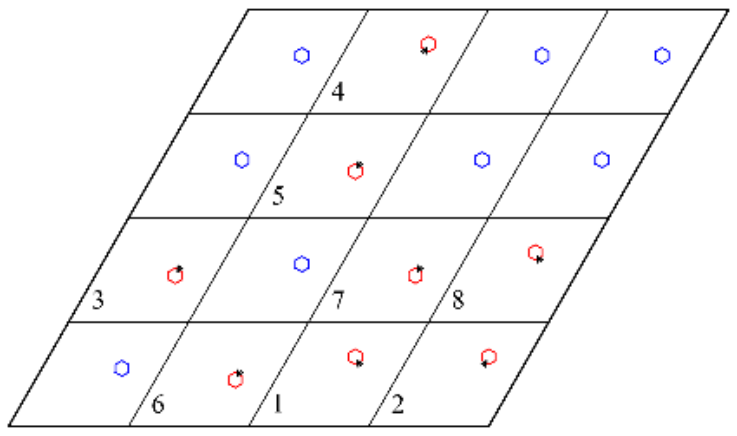 | 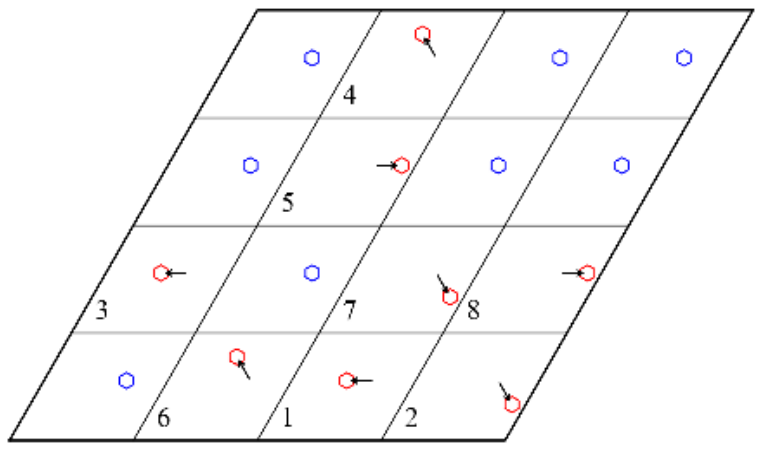 | 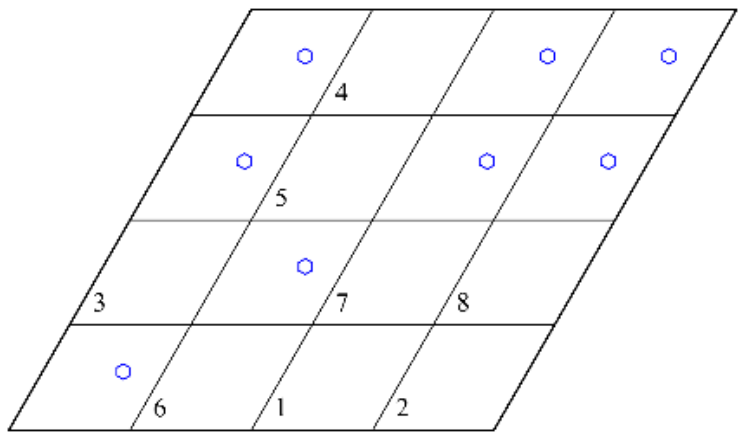 |

Configuration 1

Configuration 2

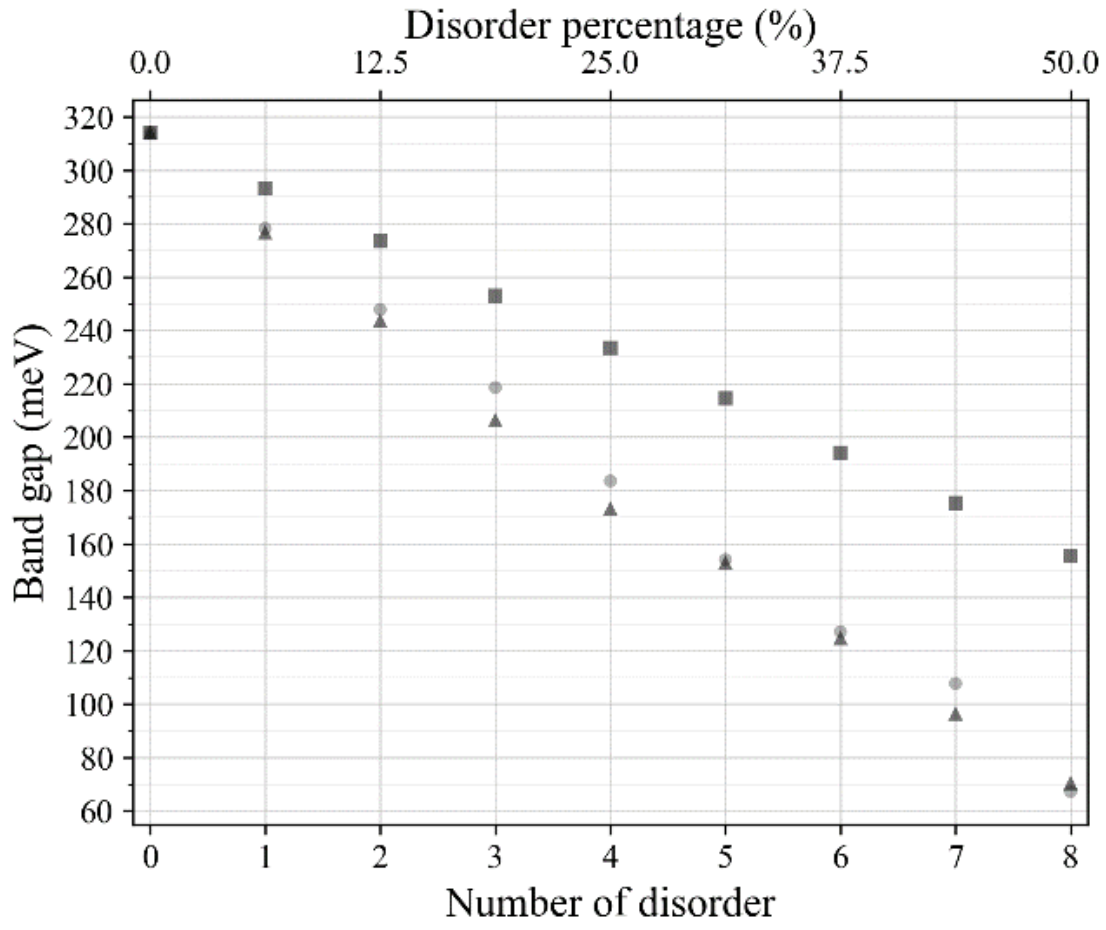

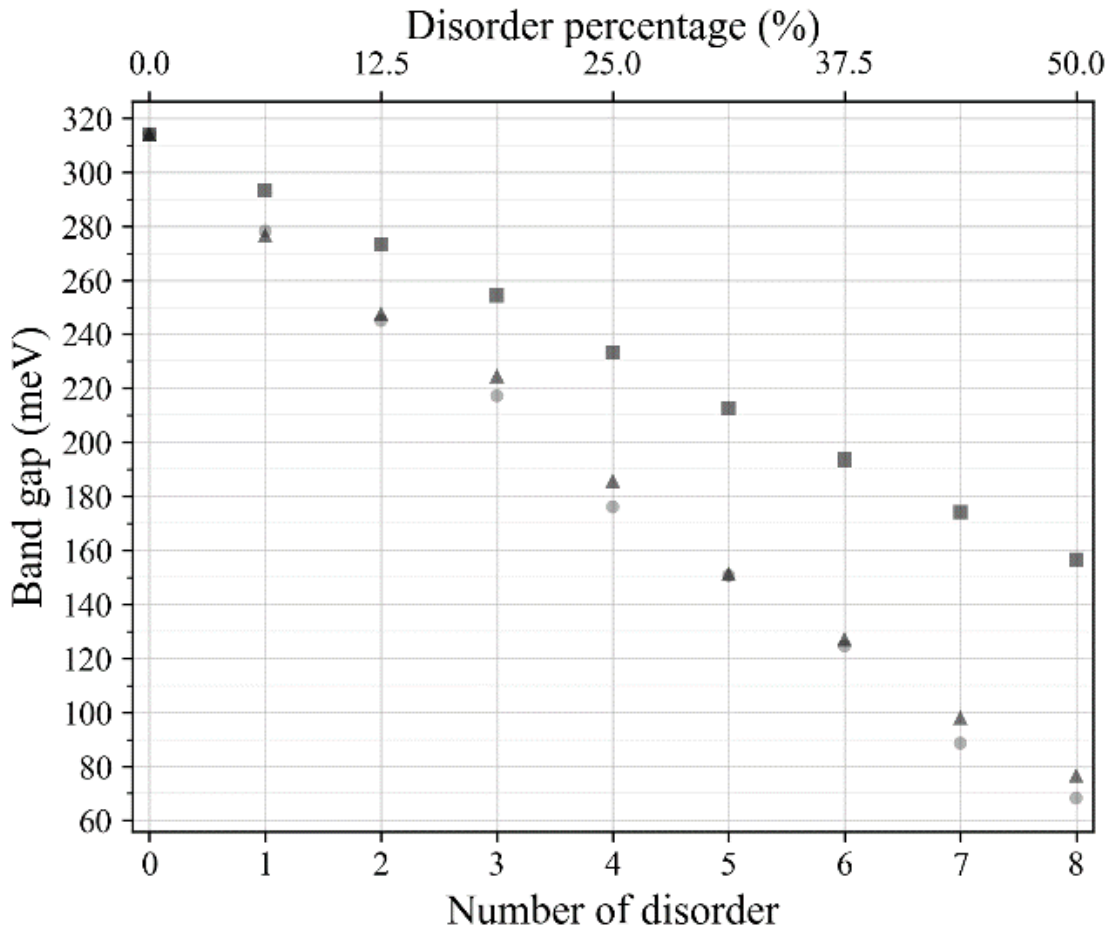

Configuration 3

Configuration 4

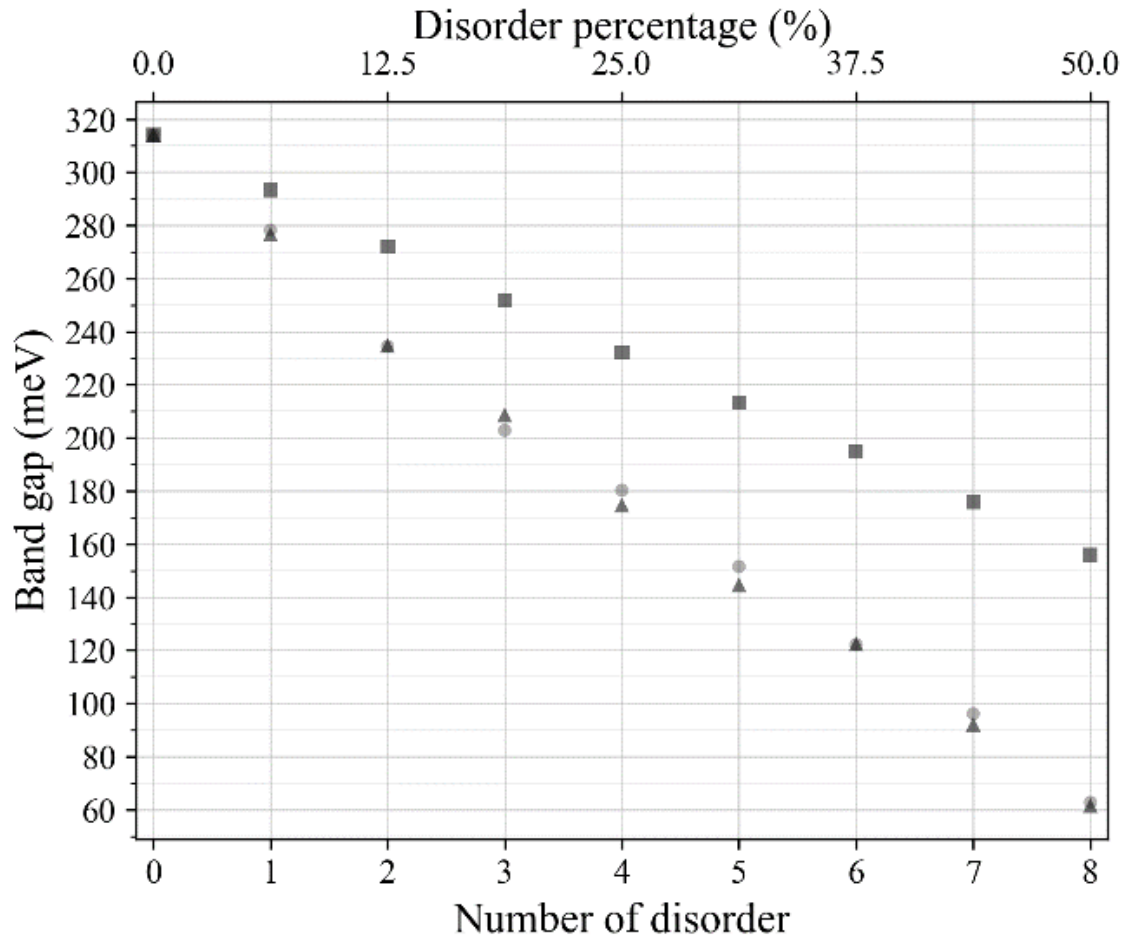

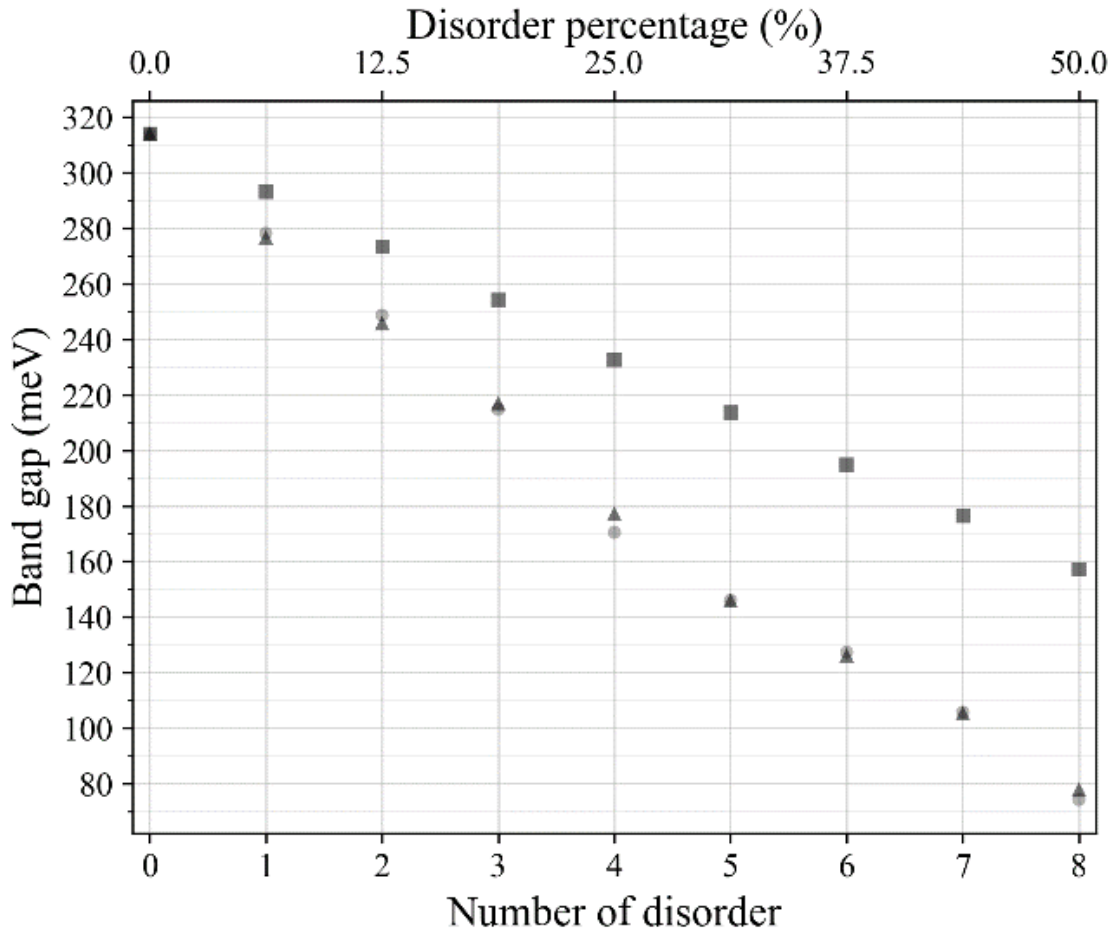

Configuration 5

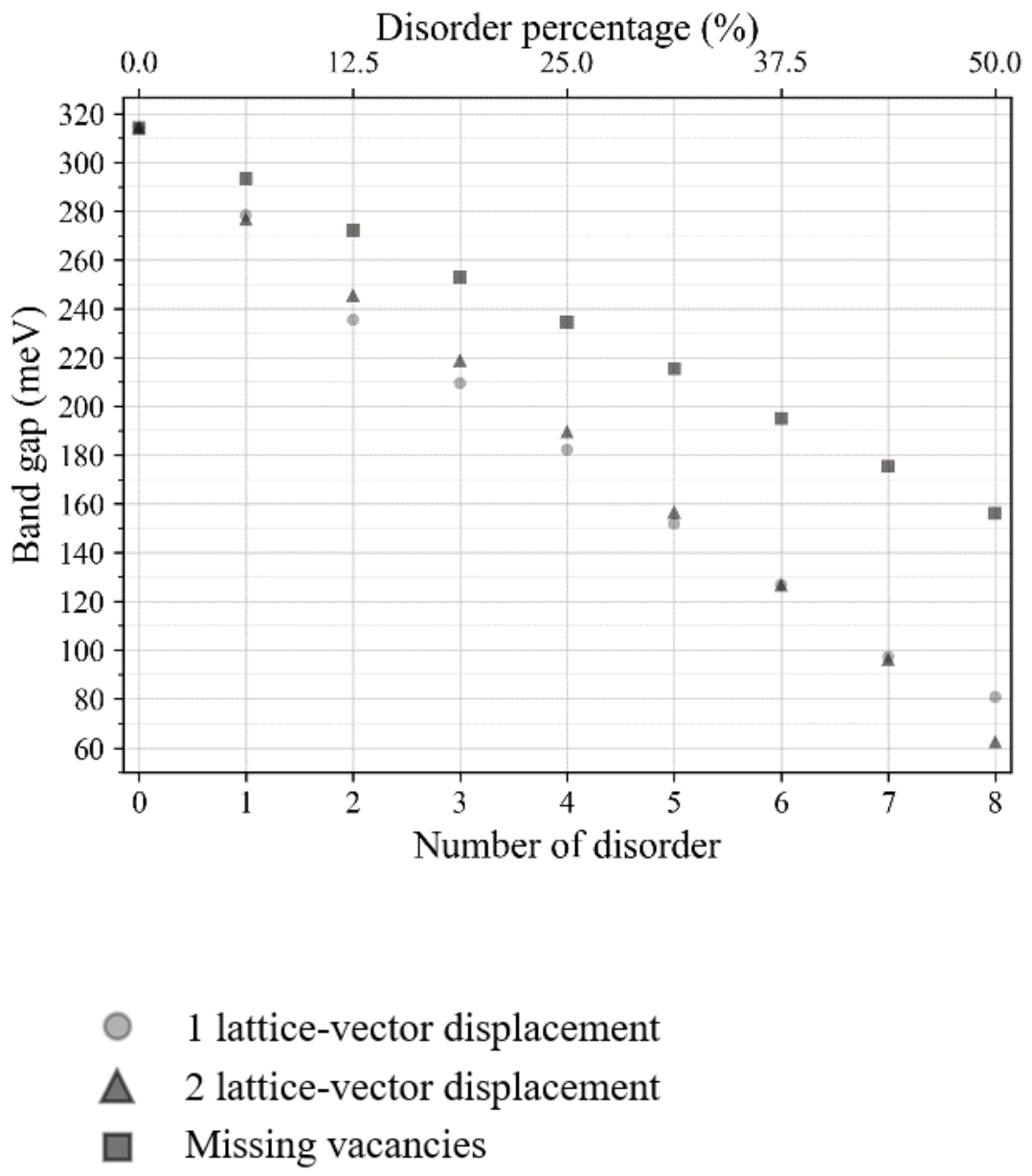

FIG. S3 Effect of disorder on the band gap in configurations 1 to 5.

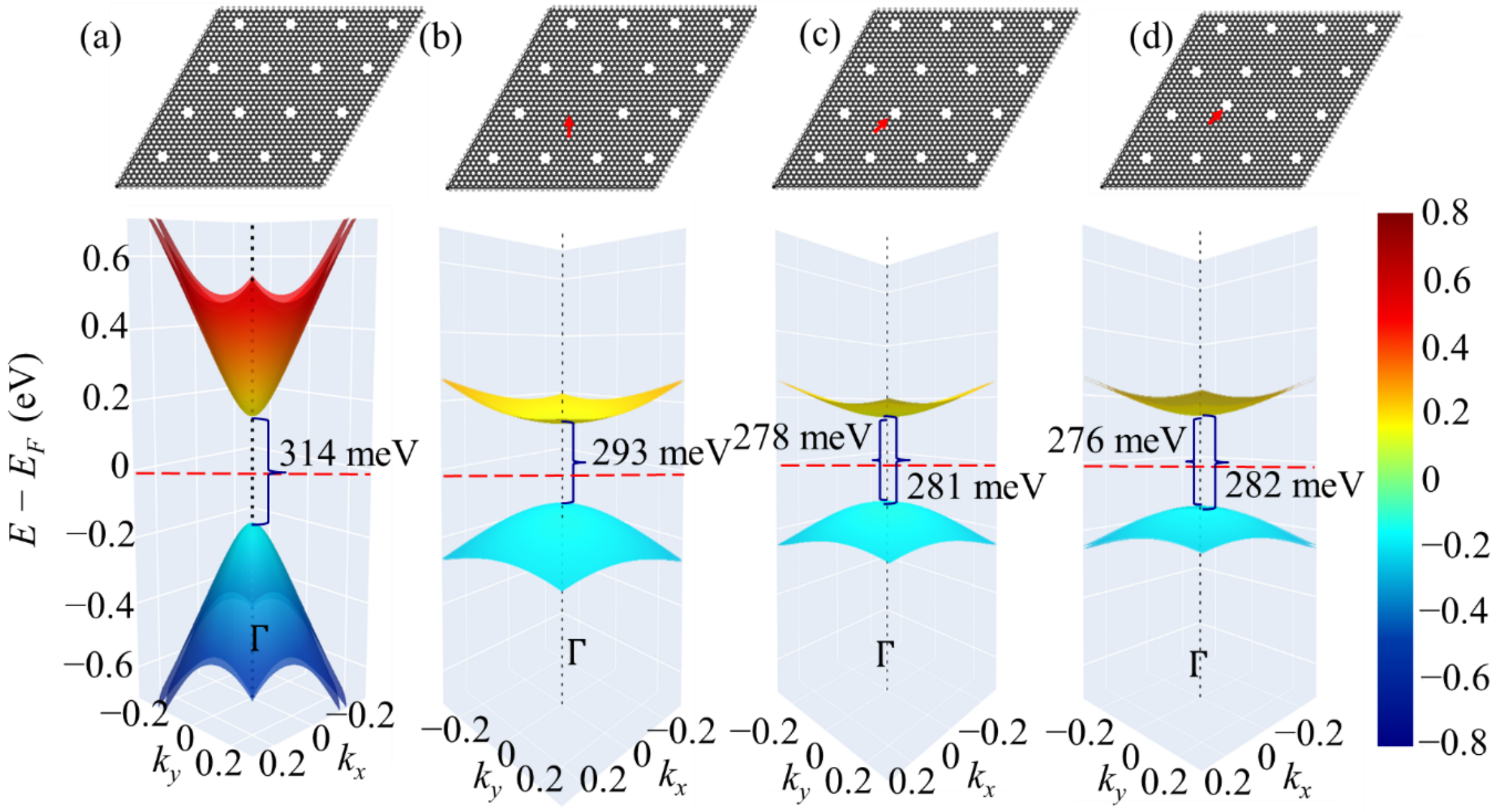

FIG. S4 Band gap in (a) ordered enlarged GSL containing 16 $D_{6h(6C)}$ vacancies, (b) with one missing vacancy, (c), (d) one displacement translated by one or two primitive lattice vectors of graphene, with the direction randomly selected from $\pm\mathbf{a_1}$, $\pm\mathbf{a_2}$, and $\pm(\mathbf{a_2} - \mathbf{a_1})$, respectively.

Table V. Linear fits of the band-gap reduction for different disorder models.

| Disorder | Linear equation | $R^2$ |
|---|---|---|
| 1 lattice-vector displacement | $E_{gap}(x) = -30.10x + 306.16$ | 0.99 |
| 2 lattice-vector displacement | $E_{gap}(x) = -30.25x + 306.87$ | 0.99 |
| Missing vacancies | $E_{gap}(x) = -19.68x + 312.77$ | 0.99 |